\documentclass[lettersize,journal]{IEEEtran}
\usepackage{algorithmic}
\usepackage{algorithm}
\usepackage{array}
\usepackage[caption=false,font=normalsize,labelfont=sf,textfont=sf]{subfig}
\usepackage{textcomp}
\usepackage{stfloats}
\usepackage{url}
\usepackage{verbatim}
\usepackage{graphicx}
\usepackage{CJKutf8} 
\usepackage{booktabs}   
\usepackage{setspace} 
\usepackage{array} 
\usepackage{subeqnarray}
\usepackage{cite}
\usepackage{amsmath,amsthm,amssymb,amsfonts} 
\usepackage{cases}
\usepackage{hyperref}
\usepackage{float}   
\usepackage{bm}
\usepackage{color}

\begin{document}
\begin{CJK}{UTF8}{gbsn} 
\title{Cooperative Base Station Assignment and Resource Allocation for 6G ISAC Network}

\author{ Jiajia Liao, \emph{Student Member, IEEE}, Luping Xiang, \emph{Senior Member, IEEE}, Shida Zhong, \emph{ Member, IEEE}, Lixia Xiao, \emph{Member, IEEE}, Haochen Liu, \emph{ Member, IEEE}, Kun Yang, \emph{Fellow, IEEE}

\thanks{This work was funded in part by the Natural Science Foundation of China under Grant 62132004 and Grant 62301122, in part by Jiangsu Major Project on Fundamental Research under Grant BK20243059, in part by Gusu Innovation Project under Grant ZXL2024360, and in part by the High-Tech District of Suzhou City under Grant RC2025001. (\textit{Corresponding author: Luping Xiang.})}
\thanks{Jiajia Liao is with the School of Information and Communication Engineering, University of Electronic Science and Technology of China, Chengdu 611731, China (e-mail: 202321010921@std.uestc.edu.cn). }
\thanks{Luping Xiang and Kun Yang are with the State Key Laboratory of Novel Software Technology, Nanjing University, Nanjing 210008, China, and School of Intelligent Software and Engineering, Nanjing University (Suzhou Campus), Suzhou 215163, China, (e-mail: luping.xiang@nju.edu.cn; kunyang@nju.edu.cn).}
\thanks{Shida Zhong is with the College of Electronics and Information Engineering, Shenzhen University, Shenzhen 518060, China (e-mail: shida.zhong@szu.edu.cn).}
\thanks{Lixia Xiao is with the Wuhan National Laboratory for Optoelectronics and the Research Center of 6G Mobile Communications, Huazhong University of Science and Technology, Wuhan 430074, China (e-mail: lixiaxiao@hust.edu.cn).}
\thanks{Haochen Liu is  with the School of Electronics and Information, Northwestern Polytechnical University, Xi'an 710019, China (e-mail: haochenliu@nwpu.edu.cn).}
}

\maketitle

\begin{abstract}
In the upcoming 6G networks, integrated sensing and communications (ISAC) will be able to provide a performance boost in both perception and wireless connectivity. This paper considers a multiple base station (BS) architecture to support the comprehensive services of data transmission and multi-target sensing. In this context, a cooperative BS assignment and resource allocation (CBARA) strategy is proposed in this paper, aiming at jointly optimizing the communication and sensing (C\&S) performance. The posterior Cramér-Rao lower bound and the achievable rate with respect to transmit power and bandwidth are derived and utilized as optimization criteria for the CBARA scheme. We develop a heuristic alternating optimization algorithm to obtain an effective sub-optimal solution for the non-convex optimization problem caused by multiple coupled variables. Numerical results show the effectiveness of the proposed solution, which achieves a performance improvement of $117\%$ in communication rate and $40\%$ in sensing accuracy, compared to the classic scheme.
\end{abstract}


\section{Introduction}
\IEEEPARstart{I}{ntegrated} sensing and communication (ISAC) has recently been recognized as a pivotal technology for the forthcoming evolution of 5G-Advanced and 6G networks \cite{lu2024integrated}, \cite{kaushik2024toward}. This innovative approach boasts numerous benefits over traditional single-function communication or standalone sensing wireless networks. By enabling the sharing of resources such as spectrum, antennas, and transceivers between communication and sensing (C\&S), ISAC enhances resource utilization efficiency, thereby reducing both capital and operational expenditures. Moreover, the inherent synergy between C\&S within ISAC frameworks can mutually enhance their performance, leading to improved overall system capabilities \cite{ISAC, xiang2023green}. 
Given these advantages, ISAC aims to offer more effective and intelligent solutions across a variety of applications, including transportation systems, autonomous vehicles, industrial automation, and smart residential environments \cite{Applications2020joint,autonomous2020joint, liu2022communication}.

While ISAC promises synergistic gains, its inherent characteristic of tightly coupled resource sharing introduces unique challenges \cite{DongSensing2024}.
Specifically, in the context of communication systems, transmit bandwidth and power affect the performance of communication systems, such as achievable rate, bit error rate and so on \cite{kivanc2003computationally,tachwali2013multiuser,zhang2017downlink,jiang2019joint}.
Similarly, in the radar field, metrics of sensing accuracy, including the Cramér-Rao lower bound (CRLB) and the posterior Cramér-Rao lower bound (PCRLB), also depend on system power and available bandwidth \cite{MIMORadarLocalization,yan2015simultaneous,zhang2020power}.  
However, C\&S share the limited bandwidth and power resources in ISAC systems. Therefore, an effective resource allocation strategy should be developed to satisfy the performance requirements of both sides, leading to improved resource utilization \cite{Luong2021RadioResource}. For instance, \cite{zhou2018resource} introduced a wireless-powered ISAC system, which jointly optimizes energy beamforming vectors, energy harvesting time, and transmit power to minimize energy consumption. In \cite{ahmed2024distributed}, a distributed dual-function radar communication multiple input multiple output (MIMO) system was proposed, employing an optimal power allocation strategy to simultaneously fulfill accurate localization and high-rate communication. \cite{xiang2024robust} proposed a power allocation solution to maximize the achievable rate under both perfect and imperfect channel estimation, based on a novel nonorthogonal multiple access (NOMA)-assisted orthogonal time-frequency space (OTFS)- ISAC network. Additionally, \cite{9729765} explored multi-user terminal scenarios integrated with mobile edge computing (MEC), which aimed to optimize the radar beampattern and computation offloading energy consumption to satisfy both C\&S demands. These studies provide reference solutions for resource allocation to trade-off communication and sensing performance for ISAC systems in different scenarios. 

Given the ease of analysis and typicality of single BS scenarios, there have been several studies focusing on resource allocation for single BS ISAC systems. For example, \cite{dong2022sensing} proposed a unified ISAC resource allocation framework for detection, localization and tracking in single BS with multiple targets network. \cite{li2024maximizing} proposed a value of service oriented resource allocation scheme to guarantee fairness among all users in the collaborative ISAC system. However, there are certain performance bottlenecks in single BS scenarios, such as limited angular resolution and sensing accuracy, restricted coverage capacity, insufficient capability for tracking mobile targets, which can be overcome by extending to multiple BSs scenarios.
 
 Compared to single BS systems, multiple BSs systems increase the coverage area and sensing accuracy in radar field \cite{yan2018power}, thus enabling ISAC systems to achieve stronger environmental awareness and enhanced communication quality \cite{wei2023integrated,meng2024cooperative,wei2024deepCooperation}. In radar field, a decentralized network architecture is considered in \cite{tharmarasa2011decentralized,xie2017joint,yi2020resource}, which optimized battery life times, transmit power and beam scheduling, respectively. Additionally, \cite{chen2015cooperative} conducted a coordinated optimization of target-radar assignment and dwell time. 
 In the context of ISAC, \cite{meng2024cooperativeISAC} introduced new cooperation regimes in network-level and discussed a range of cooperative C\&S architectures including at the task, data, and signal levels. 
\cite{XiaSymbiotic} proposed a bistatic ISAC system model, which achieved optimization of beamforming through mutual assistance between communication and sensing.
 \cite{Meng2023Performance} and \cite{Coordinated2022} explored the influence of power allocation strategies on the performance of cooperative BSs. In \cite{zhang2023joint}, a strategy of joint sub-band allocation, user association, and transmission power control was proposed to address interference management in ISAC dense cellular networks. Moreover, resource allocation problems in multistatic ISAC systems have been extensively explored across a variety of scenarios, such as the MEC-enabled ISAC networks \cite{LiuOffloading}, which jointly optimized beamforming and offloading design, the ISAC-aided \textit{ad hoc} networks \cite{JiaxingAd-Hoc}, where a resource allocation scheme based on graph theory was proposed, and the unmanned aerial vehicles (UAVs) ISAC network where UAVs work as subsidiary BSs \cite{meng2022multi} and BSs \cite{pan2024cooperativeUAV} for resource allocation. Although these studies have provided numerous solutions to the ISAC resource allocation problem, they rarely addressed dynamic scenarios with decentralized BSs for multiple targets. 

Motivated by these considerations, this paper extends the existing literature by proposing a cooperative BS assignment and resource allocation (CBARA) strategy, specifically designed for multi-target sensing and communication in dynamic ISAC networks. A detailed comparison between this work and existing studies is presented in Table \ref{tab:tt1}. 
\begin{table*}[ht!]
\footnotesize
  \begin{center}
    \caption{Boldly Contrasting Our Contributions to Existing Work}
    \label{tab:tt1} 
    \begin{tabular}{l|c|c|c|c|c|c|c|c|c|c|c} 
    \hline
      Contributions       & This work & \cite{yan2015simultaneous} &
      \cite{zhang2020power}& \cite{ahmed2024distributed}&\cite{xiang2024robust}
      &\cite{dong2022sensing},\cite{li2024maximizing}
      &\cite{yan2018power} &\cite{tharmarasa2011decentralized,chen2015cooperative} & \cite{xie2017joint,yi2020resource} &\cite{Meng2023Performance,zhang2023joint,meng2022multi}&\cite{LiuOffloading,JiaxingAd-Hoc}\\
    \hline
       ISAC              &$\checkmark$ &   
       &   &$\checkmark$ &$\checkmark$&$\checkmark$ && & &$\checkmark$&$\checkmark$\\
      \hline
       Power allocation     &$\checkmark$ 
    &$\checkmark$& $\checkmark$&$\checkmark$& $\checkmark$&$\checkmark$&$\checkmark$& &$\checkmark$ &$\checkmark$& \\
      \hline
       Bandwidth allocation  &$\checkmark$  &  
       &$\checkmark$& & &$\checkmark$ & & & & & \\
      \hline
       Multi-target C\&S &$\checkmark$                               
       & $\checkmark$ &$\checkmark$ &$\checkmark$ &$\checkmark$&$\checkmark$ &  & $\checkmark$& $\checkmark$& & \\
      \hline
      Multiple BSs           &$\checkmark$ &
      &              &$\checkmark$ &  && $\checkmark$&$\checkmark$& $\checkmark$&$\checkmark$&$\checkmark$\\
      \hline
    \end{tabular}
  \end{center}
\end{table*}
\subsection{Main Contributions}
The main contributions of this paper are summarized as follows.
\begin{enumerate}
\item We propose a multi-BS ISAC network architecture designed to provide ISAC and sensing services for moving targets in the network. A CBARA strategy is developed  by considering both BS assignment and resource allocation. Mathematically, the CBARA strategy aims to optimize both system sensing accuracy and achievable data rate simultaneously, while achieving an effective trade-off between C\&S performance through strategic resource distribution.

\item Instead of employing the conventional sensing accuracy measurement CRLB, which only characterizes the static state of the sensing targets, we derive the PCRLB for multiple BSs scenarios to track the moving states. 
Meanwhile, the performance of communication is measured in terms of the system achievable rate. In this case, the proposed CBARA strategy can be formulated as minimizing the weighted sum of C\&S metrics, facilitating joint optimization of BS assignment, transmit power and bandwidth allocation, subject to system resource constraints.

\item An effective heuristic alternating optimization algorithm is developed to obtain a suboptimal solution to the proposed non-convex optimization problem. It is challenging to find an optimal solution for this optimization problem due to the binary nature of target-to-BS allocation variables and the coupling between power and bandwidth variables. Given the high complexity of directly obtaining the optimal solution, a heuristic algorithm is first proposed to determine the BS assignment. Subsequently, an alternating optimization algorithm is employed to allocate power and bandwidth resources, enabling an efficient solution to the optimization problem with a unified objective.

\item Numerical simulation results are provided to validate the superiority of our proposed CBARA strategy in multi-target sensing and communication within ISAC network. Specifically, we initially analyze the trade-off between C\&S performance, and present the results of BS assignment and resource allocation to demonstrate the effectiveness of the CBARA strategy. Subsequently, 
the superiority of the CBARA strategy is demonstrated by contrasting it with alternative strategies.
\end{enumerate}

The rest of this paper is organized as follows. The mathematical model of the proposed ISAC system is given in Section II. In Section III, the CBARA optimization problem is formulated, and the solution is presented. The simulation results and the trade-off analysis for the C\&S performance are provided in Section IV, and the conclusions are discussed in Section V. 

As for our notation, boldface characters represent matrices (upper case) or vectors (lower  case), while normal characters denote scalars. For a tow-dimensional matrix $\bf{A}$ of any size, ${\bf{A}}^T$ and ${\bf{A}}^H$ represent its transpose and conjugate transpose, respectively. For a square matrix $\mathbf{B}$, $\mathrm{tr}(\mathbf{B})$ and ${\mathbf{B}}^{-1}$ denote its trace and inverse matrix, respectively, and $\mathbf{B}\succeq \mathbf{0}$ means that $\mathbf{B}$ is a positive semidefinite matrix, where $\mathbf{0}$ represents a zero matrix with the same size as $\mathbf{B}$. $|\cdot|$ stands for the magnitude of a complex number. $\mathrm{diag}(\mathbf{a})$ denotes the diagonal matrix constructed with the elements of $\mathbf{a}$ as its diagonal elements. ${\bf{I}}_{N}$ denotes an identity matrix with dimension $N \times N$. ${\bf{1}}_M$ and ${\bf{1}}_{M\times N}$ respectively denote $M\times1$ vector and $M\times N$ matrix, both consisting entirely of ones. $\otimes$ is the Kronecker operator.  $\mathbb{C}^{M \times N}$ denotes the space of $M\times N$ complex matrices. $\mathbb{R}^{M \times N}$ denotes the space of $M\times N$ real matrices. $\mathcal{N}(\mu, \sigma)$ represents a Gaussian distribution with a mean of $\mu$ and a variance of $\sigma$.
\vspace{-0.6cm}
\begin{figure}[htbp]
\centering
\setlength{\abovecaptionskip}{0.1cm}
\includegraphics[width=0.5\textwidth]{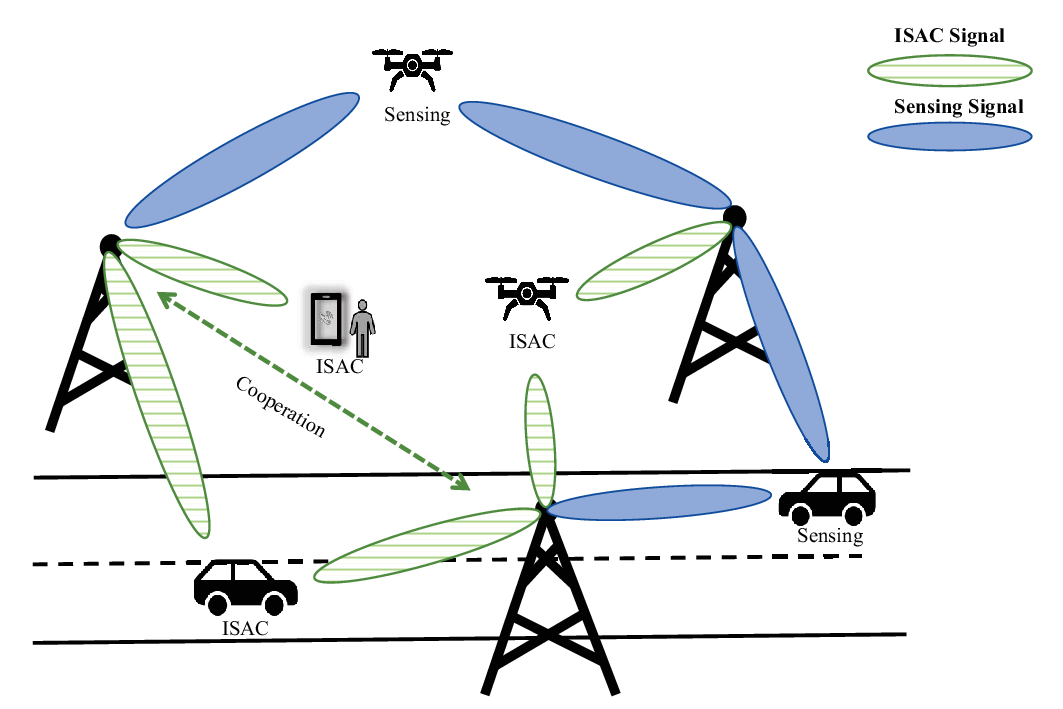}
\caption{An illustration of multi-BS ISAC network.}
\label{F1}
\end{figure}
\section{System Model}
We consider an ISAC system with $K$ MIMO BSs, as shown in Fig. \ref{F1}, where each BS is equipped with $N_t$ transmit antennas and $N_r$ receive antennas, transmitting radar signals and ISAC signals simultaneously to $Q$ sensing targets and $I$ ISAC users. 
During a mission period, BSs provide sensing services to sensing targets, while providing ISAC services to ISAC users, which adopt orthogonal frequency division multiple access (OFDMA) \footnote{The subsequent signal models are formulated on a per-subcarrier basis. Each user may be allocated multiple narrowband subcarriers for data transmission that allow the assumption of frequency-flat fading within each subcarrier.}. The mission period is discretized into $N$ time slots with fixed duration $T_s$, which represents the resource allocation update interval and the measurement sampling interval.
The BSs are indexed by the set $\mathcal{K}=\{1, \ldots, k, \ldots, K\}$, the targets by $\mathcal{Q}=\{1, \ldots, q, \ldots, Q\}$, and the ISAC users by $\mathcal{I}=\{1, \ldots, i, \ldots, I\}$. Additionally, we define a set $\mathcal{M}=\{1, \ldots, m, \ldots, M\}$, where we have $M=Q+I$, to compass all ISAC users and sensing targets, addressing situations that involve both types of objects without differentiation.
The state of the $m$-th object in the $n$-th time slot is $\mathbf{x}_{m}^{n}=\left[x_{m}^{n}, y_{m}^{n}, \dot{x}_{m}^{n}, \dot{y}_{m}^{n}\right]^T$, where $\left(x_{m}^{n}, y_{m}^{n}\right)$ and $\left(\dot{x}_{m}^{n}, \dot{y}_{m}^{n}\right)$ are the position and velocity components in the Cartesian coordinate, respectively.

We define $\mathcal{S}_{\mathrm{o},m}^n$ as the set of BSs that are assigned to serve the $m$-th object in the $n$-th slot, $\mathcal{S}_{\mathrm{b},k}^n$ as the set of objects that the $k$-th BS serves, and introduce binary variable $u_{m, k}^n$ to represent the correlation between the $m$-th object and the $k$-th BS
\begin{align}
u_{m, k}^n=\left\{\begin{array}{ll}
1, & \text { if } k \in \mathcal{S}_{\mathrm{o},m}^n\ \text{or}\ m \in \mathcal{S}_{\mathrm{b},k}^n\\
0, & \text { if } k \not\in \mathcal{S}_{\mathrm{o},m}^n\ \text{and}\ m \not\in \mathcal{S}_{\mathrm{b},k}^n 
\end{array}.\right.
\end{align}
In addition, we provide the following assumptions to simplify the problem.

\textit{Assumption 1:} The number of targets and users, as well as their initial states, have been obtained through radar detection and localization techniques \cite{yan2015simultaneous}. And the targets and the users are randomly distributed in the BS service area.

\textit{Assumption 2:} 
Each BS receives only its own echo signals, preventing interference from other radar echoes \cite{xie2017joint}.

\textit{Assumption 3:} 
To avoid interference, we employ a non-overlapping bandwidth allocation scheme for all C\&S objects \cite{dong2022sensing}.
\subsection{Signal Model}
The transmitted signal of $k$-th BS in the $n$-th slot $\mathbf{s}^{n}_{k}(t)$ is defined as 
\begin{equation}
\mathbf{s}^{n}_{k}(t)=\mathbf{F}^{n}_{k} \hat{\mathbf{s}}^{n}_{k}(t),
\end{equation}
where $\hat{\mathbf{s}}^{n}_{k}(t)=\left[\hat{s}_{1, k}^{n}(t), \ldots, \hat{s}_{M^{'}, k}^{n}(t)\right]^{T} \in \mathbb{C}^{M^{'} \times 1}$ is the baseband transmitted signal, $\mathbf{F}^{n}_{k}=\left[\mathbf{f}_{1, k}^{n}, \cdots, \mathbf{f}_{M^{'}, k}^{n}\right] \in \mathbb{C}^{N_{t} \times M^{'}}$ is the transmitted beamforming matrix, and $\mathbf{f}_{m, k}^{n}$ is the transmitted beamforming vector from the $k$-th BS to the $m$-th object. $M^{'}$ is the total number of objects served by the $k$-th BS and $m \in \mathcal{S}_{\mathrm{b},k}^n$.
 
 Each BS utilizes a uniform linear array (ULA) with a spacing equal to half the wavelength. Thus, we can obtain the following transmit and receive steering vectors
\begin{align}
\mathbf{a}_t(\theta)=&\sqrt{\frac{1}{N_t}}\left[1, e^{j \pi \sin \theta}, \cdots, e^{j \pi\left(N_t-1\right) \sin \theta}\right]^T, \\
\label{a_r}\mathbf{a}_r(\theta)=&\sqrt{\frac{1}{N_r}}\left[1, e^{j \pi \sin \theta}, \cdots, e^{j \pi\left(N_r-1\right) \sin \theta}\right]^T .
\end{align}

The received echo signal model of the $k$-th BS in the $n$-th slot is given by \cite{dong2022sensing} as
\begin{align}
\mathbf{r}^{n}_{k}= &\sum_{m \in \mathcal{S}_{\mathrm{b},k}^n} A_{m, k}^n \sqrt{p_{m, k}^n} e^{j 2 \pi f_{m, k}^n t} 
 \left(\boldsymbol{\omega}_{m, k}^n\right)^H\mathbf{a}_r\left(\theta_{m, k}^{s,n}\right) \nonumber \\
&\mathbf{a}_t^H\left(\theta_{m, k}^{s,n}\right) \mathbf{f}_{m, k}^{n}\hat{s}_{m, k}^{n}\left(t-\tau_{m, k}^n\right)+\left(\boldsymbol{\omega}_{m, k}^n\right)^H\boldsymbol{\upsilon}_{m, k}^n(t),
\end{align}
where $A_{m, k}^n$ represents the object's reflection gain, which is defined as
\begin{equation}
    A_{m, k}^n = \varrho\alpha_{m, k}^{n} h_{m, k}^{n},
\end{equation}
where $\alpha_{m, k}^{n}$ is the path-loss factor, which satisfies
$\alpha_{m, k}^{n}\propto 1/(d_{m, k}^{n})^2$, with $d_{m, k}^{n}$ being the distance between the $m$-th object and the $k$-th BS.
$\varrho=\sqrt{N_t N_r}$ and $h_{m, k}^{n}$ represent the array gain factor and the radar cross section (RCS), respectively.
$p_{m, k}^n$, $f_{m, k}^n$ and $\tau_{m, k}^n$ are the transmit power, Doppler frequency, and the time-delay, respectively. $\theta_{m, k}^{s,n}$ is the angle information of the $m$-th object. 
$\boldsymbol{\omega}_{m, k}^n \in \mathbb{C}^{N_r\times1}$ is the received beamforming vector. The term $\boldsymbol{\upsilon}_{m, k}^n \in \mathbb{C}^{N_r\times1}$ represents the complex additive white Gaussian noise (AWGN) with zero mean and a variance of $\sigma_r^2$.

The received communication signal of the $i$-th ISAC user is
\begin{align}
\mathbf{z}_{i}^n(t) =&\sum_{k \in \mathcal{S}_{\mathrm{o},i}^n} A_{i, k}^{'n} \sqrt{p_{i, k}^n} e^{j 2 \pi f_{i, k}^{'n}t} \left(\boldsymbol{\omega}_{i, k}^{'n}\right)^H \mathbf{a}_u\left(\theta_{i, k}^{c,n}\right) \nonumber\\
& \mathbf{a}_t^H\left(\theta_{i, k}^{c,n}\right) \mathbf{f}_{i, k}^n \hat{s}_{i, k}^n\left(t-\tau_{i, k}^{'n}\right)+\left(\boldsymbol{\omega}_{i, k}^{'n}\right)^H\boldsymbol{\nu}_{i, k}^n(t), 
\end{align}
where $A_{i, k}^{'n}$ is the receive gain of the $i$-th user
\begin{equation}
    A_{i, k}^{'n} = \varrho^{'}\alpha_{i, k}^{'n},
\end{equation}
where ${\varrho}^{'} = \sqrt{N_t N_r^{'}}$ is the array gain factor; $\alpha_{i, k}^{'n}$ is the pass-loss factor of communication; and $N_r^{'}$ is the number of receive antennas of the $i$-th user. $\boldsymbol{\omega}_{i,k}^{'n}\in \mathbb{C}^{N_r^{'}\times1}$ is the received beamforming vector. $\mathbf{a}_u(\theta_{i, k}^{c,n})$ is the steering vector of the ISAC user, similar to \eqref{a_r} with $N_r^{'}$ antennas. Correspondingly, the terms $f_{i, k}^{'n}$, $\tau_{i, k}^{'n}$ and $\theta_{i, k}^{c,n}$ respectively represent the Doppler frequency, the time-delay and the angle information of the $i$-th user. Finally, $\boldsymbol{\nu}_{i, k}^n(t) \in \mathbb{C}^{N_r^{'}\times1}$ is AWGN with zero mean and variance $\sigma_c^2$.

To simplify the analysis, we generate the transmit and receive beamforming vectors as \cite{dong2022sensing}
\begin{equation}
\mathbf{f}_{m, k}^n=\mathbf{a}_t(\hat{\theta}_{m, k}^n), 
 \boldsymbol{\omega}_{m, k}^n=\mathbf{a}_r(\hat{\theta}_{m, k}^{s,n}), \boldsymbol{\omega}_{i, k}^{'n}=\mathbf{a}_u(\hat{\theta}_{i, k}^{c,n}),
\end{equation}
where $\hat{\theta}_{m, k}^n$ is the estimated angle-of-departure (AoD) of the signal from the $m$-th object to the $k$-th BS. Correspondingly, the estimated angles-of-arrival (AoAs) for the sensing objects and ISAC users are defined as $\hat{\theta}_{m, k}^{s,n}$ and $\hat{\theta}_{i, k}^{c,n}$, respectively.

\subsection{Communication Rate}
In this work, we use achievable rate to evaluate communication performance because of its direct dependence on power and bandwidth. Firstly, the received SNR of the $i$-th ISAC user communicating with the $k$-th BS in the $n$-th slot is given by
\begin{equation}\label{SNR}
\begin{aligned}
 \gamma_{i, k}^n= \frac{p_{i, k}^n\left|A_{i, k}^{'n} (\boldsymbol{\omega}_{i,k}^{'n})^H \mathbf{a}_u(\theta_{i, k}^{c,n}) \mathbf{a}_t^H(\theta_{i, k}^{c,n}) \mathbf{f}_{i, k}^n\right|^2}{b_{i, k}^n \sigma_c^2}. 
\end{aligned}
\end{equation}
The communication channel gain is denoted as
\begin{equation}
\varsigma_{i, k}^n=\left|A_{i, k}^{'n}(\boldsymbol{\omega}_{i,k}^{'n})^H \mathbf{a}_u(\theta_{i, k}^{c,n}) \mathbf{a}_t^H(\theta_{i, k}^{c,n}) \mathbf{f}_{i, k}^n\right|^2.
\end{equation}
Subsequently, the achievable rate of the $i$-th ISAC user is obtained by 
\begin{equation}\label{rate}
R_i^n\left(\mathbf{p}_i^n, \mathbf{b}_i^n\right)=\sum_{k\in\mathcal{S}_{\mathrm{o},i}^n} b_{i, k}^n \log _2\left(1+\frac{p_{i, k}^n \varsigma_{i, k}^n}{b_{i, k}^n\sigma_c^2}\right),
\end{equation}
where $b_{i, k}^n$ is the bandwidth that the $k$-th BS allocates to the $i$-th user. ${\mathbf{p}}_i^n=\left[p_{i,1}^n, \dots, p_{i,K}^n\right]^{T}$ and ${\mathbf{b}}_i^n=\left[b_{i,1}^n, \dots, b_{i,K}^n\right]^{T}$ are the power and bandwidth allocation vectors of the $i$-th user, respectively.
\subsection{Sensing Accuracy}
For simplicity, we assume that the motion of the object follows a constant velocity model as \cite{yan2015simultaneous} 
\begin{equation}\label{eq:STATE}
\mathbf{x}_{m}^n=\mathbf{F}_x \mathbf{x}_{m}^{n-1}+\mathbf{v}_{m}^{n-1},
\end{equation}
where $\mathbf{x}_{m}^n=\left[x_{m}^n, y_{m}^n, \dot{x}_{m}^n, \dot{y}_{m}^n\right]^T$ represents the state vector of the $m$-th sensing object in the $n$-th time slot, and
$\mathbf{F}_x$ is the transition matrix 
\begin{equation}
\mathbf{F}_x=\mathbf{I}_2 \otimes\left[\begin{array}{cc}
1 & T_s \\
0 & 1
\end{array}\right].
\end{equation}
The term $\mathbf{v}_{m, n-1}$ denotes the process noise, which is assumed to follow a zero-mean Gaussian distribution with a predetermined covariance 
\begin{equation}
\mathbf{Q}_m=\sigma_m \mathbf{I}_2 \otimes\left[\begin{array}{ll}
\frac{1}{3} T_s^3 & \frac{1}{2} T_s^2 \\
\frac{1}{2} T_s^2 & T_s
\end{array}\right],
\end{equation}
where $\sigma_m$ is the level of process noise for the $m$-th sensing object. Consequently, \eqref{eq:STATE} is a linear Gaussian model. 

In the $n$-th time slot, the measurement of the $k$-th BS to the $m$-th object is given as
\begin{equation}\label{measurement_y}
\mathbf{y}_{m, k}^n=\mathbf{g}^n_k\left(\mathbf{x}_m^n\right)+\delta_t\mathbf{v}_{m, k}^{'n}
\end{equation}
where $\mathbf{g}^n_k(\cdot)$ is a nonlinear operator that can be expressed as $\mathbf{g}^n_k(\mathbf{x}_m^n)=\left[d_{m, k}^n, v_{m, k}^n, \theta_{m, k}^{s,n}\right]^T$ with 
\begin{equation}
\left\{\begin{array}{l}
d_{m, k}^n=\sqrt{(x_{m, k}^n-x_k)^2+(y_{m, k}^n-y_k)^2}, \\
v_{m, k}^n=(\dot{x}_{m, k}^n, \dot{y}_{m, k}^n)\left(\begin{array}{c}x_{m, k}^n-x_k\\ y_{m, k}^n-y_k\end{array}\right) / d_{m, k}^n, \\
\theta_{m, k}^{s,n}=\arctan \left[(y_{m, k}^n-y_k) / (x_{m, k}^n-x_k)\right],
\end{array}\right.
\end{equation}
where $d_{m, k}^n$, $v_{m, k}^n$ and $\theta_{m, k}^{s,n}$ are the range, velocity and azimuth angle of the $m$-th object relative to the $k$-th BS, respectively. $\delta_t$ is the error coefficient caused by delay synchronization errors between the BSs which sense the same object. The term $\mathbf{v}_{m, k}^{'n}$ in \eqref{measurement_y} represents the measurement noise, which is assumed to be a zero-mean Gaussian process with the covariance matrix
\begin{equation}
\mathbf{\Sigma}_{m, k}^n=\operatorname{diag}\left(\sigma_{d_{m, k}^n}^2, \sigma_{v_{m, k}^n}^2, \sigma_{\theta_{m, k}^{s,n}}^2\right),
\end{equation}
where $\sigma_{d_{m, k}^n}^2$, $\sigma_{v_{m, k}^n}^2$ and $\sigma_{\theta_{m, k}^{s,n}}^2$ are the CRLBs of the range, velocity, and AoA estimations, respectively. Specifically, the CRLBs can be generally expressed as \cite{yan2015simultaneous}
\begin{equation}\label{CRB}
\left\{\begin{array}{l}
\operatorname{CRLB}\left(d_{m, k}^n\right) \propto\left(p_{m, k}^n\left|\zeta_{m, k}^n\right|^2 \left|B_{\mathrm{e},m,k}^n\right|^2\right)^{-1}, \\
\operatorname{CRLB}\left(v_{m, k}^n\right) \propto\left(p_{m, k}^n\left|\zeta_{m, k}^n\right|^2 \left|T_{\mathrm{e},m,k }^n\right|^2\right)^{-1}, \\
\operatorname{CRLB}\left(\theta_{m, k}^{s,n}\right) \propto\left(p_{m, k}^n\left|\zeta_{m, k}^n\right|^2 / \mathrm{W}_{\mathrm{NN}}\right)^{-1},
\end{array}\right.
\end{equation}
where $\mathrm{W}_{\mathrm{NN}}$ is the null-to-null beam width of the receive antenna \cite{yan2015simultaneous}. $\zeta_{m, k}^n$ represents the normalized sensing channel gain, which is defined in \cite{dong2022sensing} as
\begin{equation}
    \zeta_{m, k}^n=\frac{\left|A_{m,k}^n\kappa\epsilon_{m, k}^n\epsilon_{m, k}^{'n}\right|^2}{N_r\sigma_r^2},
\end{equation}
where $\kappa=\sqrt{N_tN_r}$ is the beamforming gain. $\epsilon_{m, k}^n=\left|\mathbf{a}_t(\theta_{m, k}^{s,n}) \mathbf{a}_t^H(\hat{\theta}_{m, k}^n)\right|$ and $\epsilon_{m, k}^{'n}=\left|\mathbf{a}_r(\theta_{m, k}^{s,n}) \mathbf{a}_r^H(\hat{\theta}_{m, k}^n)\right|$ represent the transmit and receive beamforming gain factor, respectively. In this paper, we assume that $\epsilon_{m, k}^n=\epsilon_{m, k}^{'n}=1$ for perfect beamforming. 
 $T_{\mathrm{e},m,k }^n$ is the effective time duration, and $B_{\mathrm{e},m,k}^n$ is the effective bandwidth, which satisfies  
\begin{equation}\label{Be}
\begin{aligned}
 \left|B_{\mathrm{e},m,k}^n\right|^2=\frac{\int_{B^n} f^2\left|\hat{S}_{m, k}^n(f)\right|^2 \mathrm{~d}f}{\int_{B^n} \left|\hat{S}_{m, k}^n(f)\right|^2 \mathrm{~d} f},
\end{aligned}
\end{equation}
where $\hat{S}_{m, k}^n(f)$ represents the frequency domain form of $\hat{s}_{m, k}^n(t)$.

Noting that the sensing accuracy depends on the effective bandwidth $B_{\mathrm{e},m,k}^n$, but communication performance is related to the signal bandwidth $B^n$. Thus, we will discuss a unified form of bandwidth resource. We assume that the time domain waveform of the transmit signal is a rectangular pulse with a duration of $T_\mathrm{p}$ and can be expressed as $\hat{S}_{m, k}^n(f)=\mathrm{sin}\pi f T_{\mathrm{p}}/\pi f$ with a finite bandwidth $B^n$. According to \eqref{Be}, $\left|B_{\mathrm{e},m,k}^n\right|^2$ can be reasonably approximated to \cite{skolnik1960theoretical,dong2022sensing}
\begin{equation}
\left|B_{\mathrm{e},m,k}^n\right|^2 \approx B^n /\left(2 \pi^2 T_{\mathrm{p}}\right).
\end{equation}
Consequently, a convenient form for the CRLBs of \eqref{CRB} can be derived as 
\begin{equation}\label{CRB2}
\left\{\begin{array}{l}
\operatorname{CRLB}\left(d_{m, k}^n\right)=\beta_1/\left(p_{m, k}^n\left|\zeta_{m, k}^n\right|^2 b_{m, k}^n\right),\\
\operatorname{CRLB}\left(v_{m, k}^n\right)=\beta_2/\left(p_{m, k}^n\left|\zeta_{m, k}^n\right|^2\right), \\
\operatorname{CRLB}\left(\theta_{m, k}^{s,n}\right)=\beta_3/\left(p_{m, k}^n\left|\zeta_{m, k}^n\right|^2\right). 
\end{array}\right.
\end{equation}
where the terms $\beta_i (i=1,2,3)$ are constants associated with system configuration, signal waveform, beamforming gain and specific signal processing algorithms \cite{liu2020radar,dong2022sensing}.

The CRLB is suitable for static scenarios since it provides a lower bound on the variance of static parameter given the measured data \cite{CRB}. The PCRLB, on the other hand, combines measured data and dynamic model, taking into account not only the measurement noise but also the system process noise, thus offering more accurate evaluation performance in dynamic scenarios \cite{Posterior_1998}. The following is the derivation of the PCRLB for multiple BSs.
Let $\hat{\mathbf{x}}_m^n$ be an unbiased estimate of $\mathbf{x}_m^n$. PCRLB shows the MSE lower bound of any estimator, which is given as \cite{Posterior_1998}
\begin{equation}
\mathbb{E}_{\mathbf{x}_m^n \mathbf{y}_m^n}\left\{\left(\hat{\mathbf{x}}_m^n-\mathbf{x}_m^n\right)\left(\hat{\mathbf{x}}_m^n-\mathbf{x}_m^n\right)^T\right\} \succeq \mathbf{J}^{-1}\left(\mathbf{x}_m^n\right),
\end{equation}
where $\mathbf{y}_m^n$ is the measurement of BSs to the $m$-th object and $\mathbb{E}_{\mathbf{x}_m^n \mathbf{y}_m^n}$ is the expectation operation taken with respect to the state and the measurement. 
$\mathbf{J}\left(\mathbf{x}_m^n\right)$ is the posterior Fisher information matrix (FIM) 
\begin{equation}\label{J1}
\mathbf{J}\left(\mathbf{x}_m^n\right)=-\mathbb{E}_{\mathbf{x}_m^n, \mathbf{y}_m^n}\left[\Delta_{\mathbf{x}_m^n}^{\mathbf{x}_m^n} \ln p\left(\mathbf{y}_m^n, \mathbf{x}_m^n\right)\right],
\end{equation}
where the operators $\Delta_\Theta^\Phi=\Delta_\Theta \Delta_\Phi^T$ and $\Delta_\Theta$ represent the second-order and first-order partial derivative operations, respectively. $p\left(\mathbf{y}_m^n, \mathbf{x}_m^n\right)$ is the joint probability density function (PDF) of $\left(\mathbf{y}_m^n, \mathbf{x}_m^n\right)$ expressed as 
\begin{equation}\label{JPDF}
p\left(\mathbf{y}_m^n, \mathbf{x}_m^n\right)=p\left(\mathbf{x}_m^n\right) p\left(\mathbf{y}_m^n \mid \mathbf{x}_m^n\right),
\end{equation}
where $p\left(\mathbf{x}_m^n\right)$ is the prior PDF of the $m$-th object's state, and $p\left(\mathbf{y}_m^n \mid \mathbf{x}_m^n\right)$ is the global likelihood probability density function (LPDF). Since the observation of each BS is independent, the global LPDF can be formulated as
\begin{equation}\label{eq:g_lpdf}
p\left(\mathbf{y}_m^n \mid \mathbf{x}_m^n\right)=\prod_{k \in \mathcal{S}_{\mathrm{o},m}^n} p\left(\mathbf{y}_{m, k}^n \mid \mathbf{x}_m^n\right),
\end{equation}
where $p\left(\mathbf{y}_{m, k}^n \mid \mathbf{x}_m^n\right)$ is the local LPDF of the $k$-th BS, which is typically defined as a Gaussian distribution
\begin{equation}\label{gaussian_g_lpdf}
p\left(\mathbf{y}_{m, k}^n \mid \mathbf{x}_m^n\right)=\mathcal{N}\left(\mathbf{g}_k^n\left(\mathbf{x}_m^n\right), \boldsymbol{\Lambda}_{m, k}^n\right),
\end{equation}
where $\mathbf{g}_k^n(\mathbf{x}_{m,k}^n)$ and $\boldsymbol{\Lambda}_{m, k}^n = \delta_t\mathbf{\Sigma}_{m, k}^n$ are the mean and covariance matrix, respectively.

According to \eqref{J1} and \eqref{JPDF}, an elegant tow-part form of posterior FIM for the convenience of calculation is formulated as
\begin{equation}
\mathbf{J}\left(\mathbf{x}_m^n\right)=\mathbf{J}_P\left(\mathbf{x}_m^n\right)+\mathbf{J}_Y\left(\mathbf{x}_m^n\right),
\end{equation}
where $\mathbf{J}_P\left(\mathbf{x}_m^n\right)$ and $\mathbf{J}_Y\left(\mathbf{x}_m^n\right)$ are the FIM of the prior information and the data FIM, respectively, and can be expressed as
\begin{subequations}\label{JPandJY}
\begin{numcases}{}
\begin{aligned}
&\mathbf{J}_P\left(\mathbf{x}_m^n\right)  = -\mathbb{E}_{\mathbf{x}_m^n}\left[\Delta_{\mathbf{x}_m^n}^{\mathbf{x}_m^n} \ln p\left(\mathbf{x}_m^n\right)\right] \\
& = \mathbf{D}_{n-1}^{22}-\mathbf{D}_{n-1}^{21}\left(\mathbf{J}\left(\mathbf{x}_m^{n-1}\right)+\mathbf{D}_{n-1}^{11}\right)^{-1} \mathbf{D}_{n-1}^{12} 
\end{aligned}\label{JPandJY_a} \\
\mathbf{J}_Y\left(\mathbf{x}_m^n\right) = -\mathbb{E}_{\mathbf{x}_m^n, \mathbf{y}_m^n}\left[\Delta_{\mathbf{x}_m^n}^{\mathbf{x}_m^n} \ln p\left(\mathbf{y}_{m}^n \mid \mathbf{x}_m^n\right)\right],\label{JPandJY_b}
\end{numcases}
\end{subequations}
where 
\begin{equation}\label{eq:DDD}
\left\{\begin{aligned}   
\mathbf{D}_{n-1}^{11}&=-\mathbb{E}_{\mathbf{x}_m^{n-1},\mathbf{x}_m^n}\left\{\Delta_{\mathbf{x}_m^{n-1}}^{\mathbf{x}_m^{n-1}} \ln p\left(\mathbf{x}_m^n \mid \mathbf{x}_m^{n-1}\right)\right\}, \\
\mathbf{D}_{n-1}^{12}&=\left(\mathbf{D}_{n-1}^{21}\right)^T\\
&=-\mathbb{E}_{\mathbf{x}_m^{n-1}, \mathbf{x}_m^{n}}\left\{\Delta_{\mathbf{x}_m^n}^{\mathbf{x}_m^{n-1}} \ln p\left(\mathbf{x}_m^n \mid \mathbf{x}_m^{n-1}\right)\right\},\\
\mathbf{D}_{n-1}^{22}&=-\mathbb{E}_{\mathbf{x}_m^{n-1}, \mathbf{x}_m^n}\left\{\Delta_{\mathbf{x}_m^n}^{\mathbf{x}_m^n} \ln p\left(\mathbf{x}_m^n \mid \mathbf{x}_m^{n-1}\right)\right\}.
\end{aligned}\right.
\end{equation}
As the linear Gaussian model is described in \eqref{eq:STATE}, the conditional PDF $ p\left(\mathbf{x}_m^n \mid \mathbf{x}_m^{n-1}\right)$ is expressed as 
\begin{equation}
    p\left(\mathbf{x}_m^n \mid \mathbf{x}_m^{n-1}\right) = \mathcal{N}\left(\mathbf{F}_x \mathbf{x}_m^{n-1},\mathbf{Q}_m\right).
\end{equation}
Dropping out the expectation operator, \eqref{eq:DDD} can be approximated as 
\begin{equation}\label{eq:DDD2}
\left\{\begin{array}{l}
\mathbf{D}_{n-1}^{11}=\mathbf{F}_x^T\left(\mathbf{Q}_m\right)^{-1} \mathbf{F}_x, \\
\mathbf{D}_{n-1}^{12}=-\mathbf{F}_x^T\left(\mathbf{Q}_m\right)^{-1}=\left(\mathbf{D}_{n-1}^{21}\right)^T, \\
\mathbf{D}_{n-1}^{22}=\left(\mathbf{Q}_m\right)^{-1}.
\end{array}\right.
\end{equation}
Subsequently, by substituting \eqref{eq:DDD2} into \eqref{JPandJY_a} and using the matrix inversion theorem, $\mathbf{J}_P\left(\mathbf{x}_m^n\right)$ can be obtained as
\begin{equation}
\mathbf{J}_P\left(\mathbf{x}_m^n\right)=\left(\mathbf{Q}_m+\mathbf{F}_x \mathbf{J}^{-1}\left(\mathbf{x}_m^{n-1}\right) \mathbf{F}_x^{T}\right)^{-1} .
\end{equation}
$\mathbf{J}_Y\left(\mathbf{x}_m^n\right)$  is derived by substituting \eqref{eq:g_lpdf} and \eqref{gaussian_g_lpdf} into \eqref{JPandJY_b}, which can be expressed as
\begin{equation}\label{JY}
\begin{aligned}
\mathbf{J}_Y\left(\mathbf{x}_m^n\right) 
=\mathbb{E}_{\mathbf{x}_m^n}\left[\sum_{k \in \mathcal{S}_{\mathrm{o},m}^n}\left(\mathbf{H}_{m, k}^n\right)^{T}\left(\mathbf{\Lambda}_{m, k}^n\right)^{-1} \mathbf{H}_{m, k}^n\right],
\end{aligned}
\end{equation}
where $\mathbf{H}_{m, k}^n$ is the Jacobin matrix of the measurement function $\mathbf{g}_k^n(\cdot)$ for the $k$-th BS's observation to the $m$-th object, implying that $\mathbf{J}_Y\left(\mathbf{x}_m^n\right)$ depends on the dynamics of the object.

Noting that $\mathbf{J}_P\left(\mathbf{x}_m^n\right)$ can be calculated recursively from the given initial value of the posterior FIM. Moreover, power and bandwidth allocations have an impact on $\mathbf{\Lambda}_{m, k}^n$.
As expectation arithmetic is difficult to solve in practice, the posterior FIM can be approximated as \cite{yan2015simultaneous}
\begin{align}
\mathbf{J}\left(\mathbf{x}_m^n\right)  =&\left(\mathbf{Q}_m+\mathbf{F}_x \mathbf{J}^{-1}\left(\mathbf{x}_m^{n-1}\right) \mathbf{F}_x^{T}\right)^{-1}\nonumber\\ &+\left.\sum_{k \in \mathcal{S}_{\mathrm{o},m}^n}\left(\mathbf{H}_{m, k}^n\right)^{T}\left(\mathbf{\Lambda}_{m, k}^n\right)^{-1} \mathbf{H}_{m, k}^n\right|_{\mathbf{x}_m^n=\hat{\mathbf{x}}_m^{n\mid n-1}}\nonumber \\
 =&\left(\mathbf{Q}_m+\mathbf{F}_x \mathbf{J}^{-1}\left(\mathbf{x}_m^{n-1}\right) \mathbf{F}_x^{T}\right)^{-1}\nonumber\\&+\sum_{k \in \mathcal{S}_{\mathrm{o},m}^n}\left(\hat{\mathbf{H}}_{m, k}^n\right)^{T}\left(\hat{\mathbf{\Lambda}}_{m, k}^n\right)^{-1} \hat{\mathbf{H}}_{m, k}^n,
\end{align}
where $\hat{\mathbf{x}}_m^{n\mid n-1}$ represents the predicted object state based on the motion model in \eqref{eq:STATE} by using the $(n-1)$-th state information with zero process noise case \cite{Beyond_2004}.

By introducing the variables of the BS assignment and resource allocations, the posterior FIM can be rewritten as
\begin{align}\label{PFIM}
&\mathbf{J}_{\mathbf{x}_m^n}\left(\mathbf{u}_m^n,\mathbf{p}_m^n, \mathbf{b}_m^n\right)  =\left(\mathbf{Q}_m+\mathbf{F}_x \mathbf{J}^{-1}\left(\mathbf{x}_m^{n-1}\right) \mathbf{F}_x^{T}\right)^{-1}\nonumber\\ &+\sum_{}u_{m,k}^n\left(\hat{\mathbf{H}}_{m, k}^n\right)^{T}\left(\hat{\mathbf{\Lambda}}_{m, k}^n\left(p_{m, k}^n, b_{m, k}^n\right)\right)^{-1} \hat{\mathbf{H}}_{m, k}^n,
\end{align}
where $\mathbf{u}_m^n=\left[u_{m,1}, \dots, u_{m,K}\right]^T$, $\mathbf{p}_m^n=\left[p_{m,1}, \dots, p_{m,K}\right]^T$ and $\mathbf{b}_m^n=\left[b_{m,1}, \dots, b_{m,K}\right]^T$ represent the BS assignment, power allocation and bandwidth allocation vectors for the $m$-th object, respectively. Consequently, the predictive PCRLB for the $m$-th object is derived as 
\begin{equation}\label{PCRLB}
\operatorname{PCRLB}_{\mathbf{x}_m^n }=\mathbf{J}_{\mathbf{x}_m^n}^{-1}\left(\mathbf{u}_m^n,\mathbf{p}_m^n, \mathbf{b}_m^n\right) .
\end{equation}

\section{CBARA FRAMEWORK}
\subsection{Problem Formulation}
Mathematically, the BS assignment and resource management problem can be described as an optimization problem, subject to the constraints of system resources. The achievable rate and the predictive PCRLB derived in Section II can be utilized as optimization metrics for the CBARA strategy. Since the steps of the solution are the same in each time slot, the time slot superscript $n$ is temporarily omitted for notational convenience.

First, let $\mathbf{U}=\left[\mathbf{u}_1, ..., \mathbf{u}_K\right] \in \left\{0, 1\right\}^{M \times K}$, $\mathbf{P}=\left[\mathbf{p}_1, ..., \mathbf{p}_K\right]\in \mathbb{R}^{M \times K}$ and $\mathbf{B}=\left[\mathbf{b}_1, ..., \mathbf{b}_K\right]\in \mathbb{R}^{M \times K}$ denote the BS assignment, power and bandwidth allocation matrices, respectively, where $\mathbf{u}_k=\left[u_{1,k}, \dots, u_{M,k}\right]^T$, $\mathbf{p}_k=\left[p_{1,k}, \dots, p_{M,k}\right]^T$, $\mathbf{b}_k=\left[b_{1,k}, \dots, b_{M,k}\right]^T$. Since our aim is to jointly optimize the performance of C\&S, the objective function can be described as the following weighted ISAC metrics:
\begin{align}
\label{objective_function}
\mathop {\min }\limits_{{{\bf{U}},}{{\bf{P}},}{{\bf{B}}}} F\left( {{{\bf{U}}},{{\bf{P}}},{{\bf{B}}}} \right)  =& \mathop {\min }\limits_{{\bf{U},}{\bf{P},}{\bf{B}}} \left[ \eta \sum\limits_{m = 1}^M \mathrm{tr}\left( {{\bf{J}}_{{\bf{x}}_m}^{ - 1}\left( {{\bf{u}}_m,{\bf{p}}_m,{\bf{b}}_m} \right)} \right)\nonumber \right.\\
 &\left.- \left( {1 - \eta } \right)\sum\limits_{i = 1}^I {R_i\left( {{\bf{u}}_i,{\bf{p}}_i,{\bf{b}}_i} \right)} \right],
\end{align}
where $\eta \in \left(0, 1\right)$ represents a given scale factor to indicate the priority preference for communication or sensing performance, and a higher $\eta$ emphasizes accurate object tracking by minimizing PCRLB, while a lower $\eta$ enhances throughput by maximizing achievable rate. Since the competition for shared power and bandwidth resources, there is a trade-off between PCRLB and achievable rate. By introducing system resource constraints, the optimization problem for CBARA is formulated as
\begin{subequations}\label{P1}
\begin{align}
(\mathrm{P1})&:
\mathop {\min }\limits_{{{\bf{U}},}{{\bf{P}},}{{\bf{B}}}}F\left( {{{\bf{U}}},{{\bf{P}}},{{\bf{B}}}} \right) \notag\\
\mathrm{s.t.}&\begin{cases}
P_{\min } \leq p_{m,k} \leq P_{\max }, & \text{if} \enspace u_{m,k}=1, \forall m,k \\
p_{m,k}=0, & \text{if} \enspace u_{m,k}=0, \forall m,k \end{cases},\label{P1_a}\\
 & \begin{cases}
       B_{\min } \leq b_{m,k} \leq B_{\max }, &\text{if} \enspace u_{m,k}=1, \forall m,k  \\
       b_{m,k}=0, & \text{if} \enspace u_{m,k}=0, \forall m,k  
   \end{cases},\label{P1_b}\\
 & \mathbf{1}_M^T \mathbf{p}_k=P_{\text {total }}, \label{P1_c}\\
    & \mathbf{1}_M^T \mathbf{B}\mathbf{1}_K =B_{\text {total }}, \label{P1_d}\\
 & L_{\min} \le \mathbf{1}_K^T\mathbf{u}_m\le L_{\max},\label{P1_e}\\
& u_{m,k} \in \left\{0,1 \right\}, \forall m,k, \label{P1_f}
\end{align}
\end{subequations}
where $\mathbf{u}_m=\left[u_{m,1}, \dots, u_{m,K}\right]^T$. The inequalities in \eqref{P1_a} and \eqref{P1_b} imply that the power and bandwidth resources allocated to each object by a BS are confined within specific limits. The minimum values ensure adequate communication or sensing performance for each object, while the maximum values prevent over-allocation and maintain system efficiency. Constraints \eqref{P1_c} and \eqref{P1_d} represent the transmit power conservation of each BS and the total bandwidth conservation for the entire system, respectively.
Finally, inequality \eqref{P1_e} constrains the number of BSs serving the same object, given the limited resources budget.

Given that the inverse operation on matrices with variables in \eqref{objective_function} is difficult, we introduce an auxiliary matrix ${\bf{D}}_m$ to relax ${{\bf{J}}_{{\bf{x}}_m}^{ - 1}\left( {{\bf{u}}_m,{\bf{p}}_m,{\bf{b}}_m} \right)}$, which satisfies the following constraint
\begin{equation}\label{D}
\mathbf{D}_m \succeq {{\bf{J}}_{{\bf{x}}_m}^{ - 1}\left( {{\bf{u}}_m,{\bf{p}}_m,{\bf{b}}_m} \right)},
\end{equation}
which means $\mathbf{D}_m-{\bf{J}}_{{\bf{x}}_m}^{ - 1}\left( {{\bf{u}}_m,{\bf{p}}_m,{\bf{b}}_m} \right) \succeq \mathbf{0}$. Based on the property of the Schur complement, 
as ${{\bf{J}}_{{\bf{x}}_m}\left( {{\bf{u}}_m,{\bf{p}}_m,{\bf{b}}_m} \right)}^{ - 1}$ is also a positive semidefinite matrix, the inequality \eqref{D} is equivalent to
\begin{equation}\label{tansDJ}
\left[\begin{array}{cc}
\mathbf{D}_m & \mathbf{I}_4 \\
\mathbf{I}_4 & {\bf{J}}_{{\bf{x}}_m}\left( {{\bf{u}}_m,{\bf{p}}_m,{\bf{b}}_m} \right)
\end{array}\right] \succeq \mathbf{0}. 
\end{equation}
Then, problem \eqref{P1} can be converted to
\begin{subequations}\label{P1.1}
\begin{align}
(\text{P1.1}&):\mathop {\min }\limits_{{{\bf{U}}},{{\bf{P}}},{{\bf{B}}},{{\bf{D}}}}  F^{'}\left( {{{\bf{U}}},{{\bf{P}}},{{\bf{B}}},{{\bf{D}}}} \right) \notag\\
\mathrm{s.t.} &
\left[\begin{array}{cc}
\mathbf{D}_m & \mathbf{I}_4 \\
\mathbf{I}_4 & {\bf{J}}_{{\bf{x}}_m}\left( {{\bf{u}}_m,{\bf{p}}_m,{\bf{b}}_m} \right)
\end{array}\right] \succeq \mathbf{0}, \label{P1.1_a}
\\& \begin{cases}
P_{\min } \leq p_{m,k} \leq P_{\max }, &\text{if} \enspace u_{m,k}=1, \forall m,k \\
p_{m,k}=0, & \text{if} \enspace u_{m,k}=0, \forall m,k\end{cases},\label{P1.1_b}\\
   & \begin{cases}
       B_{\min } \leq b_{m,k} \leq B_{\max }, &\text{if} \enspace u_{m,k}=1, \forall m,k\\
       b_{m,k}=0, &\text{if} \enspace u_{m,k}=0, \forall m,k
   \end{cases},\label{P1.1_c} \\
& \mathbf{1}_M^T \mathbf{p}_k=P_{\text {total }},\label{P1.1_d} \\
& \mathbf{1}_M^T \mathbf{B}\mathbf{1}_K =B_{\text {total }},\label{P1.1_e}  \\ 
& L_{\min} \le \mathbf{1}_K^T\mathbf{u}_m\le L_{\max},\label{P1.1_f}\\
& u_{m,k} \in \left\{0,1 \right\}, \forall m,k, \label{P1.1_g}
\end{align}
\end{subequations}
where
\begin{align}
\mathop {\min }\limits_{{{\bf{U}}},{{\bf{P}}},{{\bf{B}}},{{\bf{D}}}} F^{'}\left( {{{\bf{U}}},{{\bf{P}}},{{\bf{B}}},{{\bf{D}}}} \right) &
 = \mathop {\min }\limits_{{{\bf{U}}},{{\bf{P}}},{{\bf{B}}},{{\bf{D}}}} \left[\eta \sum\limits_{m = 1}^M \mathrm{tr}\left( \mathbf{D}_m \right)\nonumber \right. \\&
 \left.- \left( {1 - \eta } \right)\sum\limits_{i = 1}^I {R_i\left( {{\bf{u}}_i,{\bf{p}}_i,{\bf{b}}_i} \right)}\right].
\end{align}
\subsection{Proposed Solution}
Due to the binary constraint in \eqref{P1.1_g} and the coupling of power and bandwidth variables, the problem formulated above is nonconvex, making it difficult to obtain a global optimal solution by directly solving. To solve this problem, we separate it into two steps of BS assignment and resource allocation to get a sub-optimal solution. Firstly, a heuristic algorithm is employed to obtain a rational BS assignment scheme, which avoids the non-convexity of the problem imposed by the binary variable $\mathbf{U}$. Subsequently, for a given $\mathbf{U}$, the constraints in \eqref{P1.1} are now linear, and the objective function is now convex for each of the $\mathbf{b}$ and $\mathbf{p}$. The convexity of \eqref{P1.1} is proved in the Appendix. 
Accordingly, we adopt the alternative optimization (AO) method \cite{AO} to decouple the power and bandwidth variables for resource allocation. Specifically, the steps are as follows：
\begin{enumerate}
\item{\textit{BS assignment by a heuristic algorithm}: We first assume that the total bandwidth $\textit{B}_{\mathrm{tolal}}$ is uniformly allocated to each object with $\bf{B}_{\mathrm{uni}}=\mathbf{1}_{\mathit{M\times K}}\cdot\mathit{B}_{\mathrm{total}}/\mathit{M}$ }. Then, we set the BS assignment variable as $\mathbf{U}_{\mathrm{all}}=\mathbf{1}_{Q\times K}$, meaning that all the BSs in this system are assigned to serve each object. 
For the given $\bf{B}_{\mathrm{uni}}$ and $\mathbf{U}_{\mathrm{all}}$, the optimization problem (\ref{P1.1}) on the power variable can be converted into a convex problem, given by
\begin{subequations}\label{P3}
\begin{align}
(\text{P2})&:\mathop {\min }\limits_{{{\bf{P}}},{{\bf{D}}}}  F^{'}\left( {{{\bf{P}}},{{\bf{D}}}} \right) \notag\\
\mathrm{s.t.} &
\left[\begin{array}{cc}
\mathbf{D}_m & \mathbf{I}_4 \\
\mathbf{I}_4 & {\bf{J}}_{{\bf{x}}_m}\left( {{\bf{p}}_m} \right)
\end{array}\right] \succeq \mathbf{0},\\
& P_{\min } \leq p_{m,k} \leq P_{\max }, \forall m,k, \\
& \mathbf{1}_M^T \mathbf{p}_k=P_{\text {total }}. 
\end{align}
\end{subequations}
The problem \eqref{P3} can be efficiently solved by using the off-the-shelf CVX toolbox \cite{CVX} to obtain the power allocation $\tilde{\bf{P}}$. 
Obviously, the result of power allocation $\tilde{\bf{P}}$ implies a trend that a larger power allocated by the 
$k$-th BS to the $m$-th object indicates a higher likelihood that the $k$-th BS should serve that object. Accordingly, we first employ a threshold $\varphi$. Then we collect the indices of the elements $\tilde{p}_{m,k}$ whose ratio $\tilde{p}_{m,k}/P_{\mathrm{total}}$ exceeds $\varphi$, thereby generating a sub-optimal subset of BS assignment within the constraint specified in \eqref{P1.1_f}. Finally, the elements in $\hat{\mathbf{U}}$ corresponding to the indices in the sub-optimal subset obtained above are assigned a value of 1, while all other elements are set to zero.
\item{\textit{Resource allocation by AO method}: For BS assignment $\hat{\mathbf{U}}$ obtained above and a given initial bandwidth allocation $\bf{B}_{\mathrm{uni}}$, a sub-problem can be formulated as 
\begin{subequations}\label{P4}
\begin{align}
(\text{P3})&:\mathop {\min }\limits_{{{\bf{P}}},{{\bf{D}}}}  F^{'}\left( {{{\bf{P}}},{{\bf{D}}}} \right) \notag\\
\mathrm{s.t.} &
\left[\begin{array}{cc}
\mathbf{D}_m & \mathbf{I}_4 \\
\mathbf{I}_4 & {\bf{J}}_{{\bf{x}}_m}\left( {{\bf{p}}_m} \right)
\end{array}\right] \succeq \mathbf{0},\\
& \begin{cases}
P_{\min } \leq p_{m,k} \leq P_{\max }, &\text{if} \enspace u_{m,k}=1, \forall m,k \\
p_{m,k}=0, &\text{if} \enspace u_{m,k}=0, \forall m,k
\end{cases},\\
& \mathbf{1}_M^T \mathbf{p}_k=P_{\text {total }}, 
\end{align}
\end{subequations}
which can also be solved by CVX to obtain the optimal power allocation $\mathbf{P}$. Afterwards, having $\mathbf{P}$ fixed, the  variable $\mathbf{B}$ can be updated by solving the sub-problem of
\begin{subequations}\label{P5}
\begin{align}
(\text{P4})&:\mathop {\min }\limits_{{{\bf{B}}},{{\bf{D}}}}  F^{'}\left( {{{\bf{B}}},{{\bf{D}}}} \right) \notag\\
\mathrm{s.t.} &
\left[\begin{array}{cc}
\mathbf{D}_m & \mathbf{I}_4 \\
\mathbf{I}_4 & {\bf{J}}_{{\bf{x}}_m}\left( {{\bf{b}}_m} \right)
\end{array}\right] \succeq \mathbf{0},\\
& \begin{cases}
B_{\min } \leq b_{m,k} \leq B_{\max }, &\text{if} \enspace u_{m,k}=1, \forall m,k \\
b_{m,k}=0, &\text{if} \enspace u_{m,k}=0, \forall m,k 
\end{cases},\\
& \mathbf{1}_M^T \mathbf{B}\mathbf{1}_K =B_{\text {total }}.
\end{align}
\end{subequations}
Finally, optimized resource allocation solutions $\hat{\mathbf{P}}$ and $\hat{\mathbf{B}}$ can be obtained by alternatively optimizing $\mathbf{P}$ and $\mathbf{B}$ using \eqref{P4} and \eqref{P5}, respectively, until convergence.}
\end{enumerate}

For a more intuitive understanding, a brief outline of the proposed CBARA solution is summarized in $\mathbf{Algorithm}$ 1.
\begin{table}[t!]  
\begin{center}  
    \begin{tabular*}{\hsize}{@{}@{\extracolsep{\fill}}l@{}}  
        \toprule[1.5pt]  
            \textbf{Algorithm 1:}
             The summary of CBARA solution\\
        \midrule  
            \enspace1: \ Initial: $\mathbf{B}_{\mathrm{uni}}$, $\mathbf{U}_{\mathrm{all}}$, $\varphi$, $\varepsilon$, $t=1$;\\
           \quad\quad $****$\textit{Optimizing BS assignment}$****$\\
           \enspace2: \ Solve problem (\ref{P3}) using $\mathbf{B}_{\mathrm{uni}}$ and $\mathbf{U}_{\mathrm{all}}$ to\\
           \quad\quad obtain $\tilde{\mathbf{P}}$;\\
           \enspace3: \ Generate the binary matrix $\hat{\bf{U}}$ by choosing the indexes \\
           \quad \quad of elements in $\tilde{\mathbf{P}}$ whose ratio beyond $\varphi$;\\
            \quad \quad $****$\textit{Optimizing resource allocation}$****$\\
            \enspace4: \ Solve the problem \eqref{P1.1} with obtained $\hat{\bf{U}}$ and given $\mathbf{B}_{(0)}$;\\
            \enspace5: \ $\mathbf{while}$ $\left|F^{'}\left(\mathbf{P}_{(t)}, \mathbf{B}_{(t)}\right)-F^{'}\left(\mathbf{P}_{(t-1)}, \mathbf{B}_{(t-1)}\right)\right|> \varepsilon$ $\mathbf{do}$\\
            \enspace6: \ \quad Solve (\ref{P4}) with given $\hat{\bf{U}}$ and fixed $\mathbf{B}_{(t-1)}$ to obtain $\mathbf{P}_{(t)}$; \\
            \enspace7: \ \quad Solve (\ref{P5}) with given $\hat{\bf{U}}$ and fixed $\mathbf{P}_{(t)}$ to obtain $\mathbf{B}_{(t)}$; \\ 
            \enspace8: \ \quad $t=t+1$;\\
            \enspace9: \ $\mathbf{end}$ $\mathbf{while}$\\
            10: \ $\hat{\mathbf{P}}=\mathbf{P}_{(t)}$,  $\hat{\mathbf{B}}=\mathbf{B}_{(t)}$;\\
            11: \ Output: $\hat{\bf{U}}$, $\hat{\mathbf{P}}$ and $\hat{\mathbf{B}}$.\\
       \bottomrule  
    \end{tabular*}
    \label{tab1}
    \vspace{-0.5cm}
\end{center}
\end{table}
\subsection{Computational Complexity}
The computational complexity of the proposed algorithm is primarily affected by the number of iterations required to solve each sub-problem in \eqref{P4} and \eqref{P5}. Since each sub-problem shares a similar structure, they exhibit identical computational complexity. Assuming that the solution to the resource allocation problem necessitates $l$ alternating optimizations, the maximum complexity of our algorithm would be $\mathcal{O}(2l+1)$. 
In contrast, it is evident that the exhaustive algorithm, which can indeed obtain the optimal BS assignment scheme by  thorough searching, incurs a significantly higher computational complexity of $\mathcal{O}\left(2l\sum\limits_{L=L_{\mathrm{min}}}^{L_{\mathrm{max}}}\binom{L}{K}\right)$. 
In the worst-case scenario, when $L_{\mathrm{min}}$ and $L_{\mathrm{max}}$ are sufficiently large and approach $K/2$, $\binom{L}{K}$ approaches its maximum value $\binom{K/2}{K}$. Consequently, the computational complexity of the exhaustive algorithm in this case can be approximated by $\mathcal{O}\left(2l\left(L_{\mathrm{max}}-L_{\mathrm{min}}+1\right)\binom{K/2}{K}\right)$, which increases exponentially with $K$. Therefore, our proposed algorithm has a reduced complexity compared to the exhaustion-based solution, while ensuring the effectiveness of the solution, which can be verified from the following simulation results.
\section{Simulation Results} 
\begin{figure}[h]
\centering
\setlength{\abovecaptionskip}{0.1cm}
\includegraphics[width=0.5\textwidth]{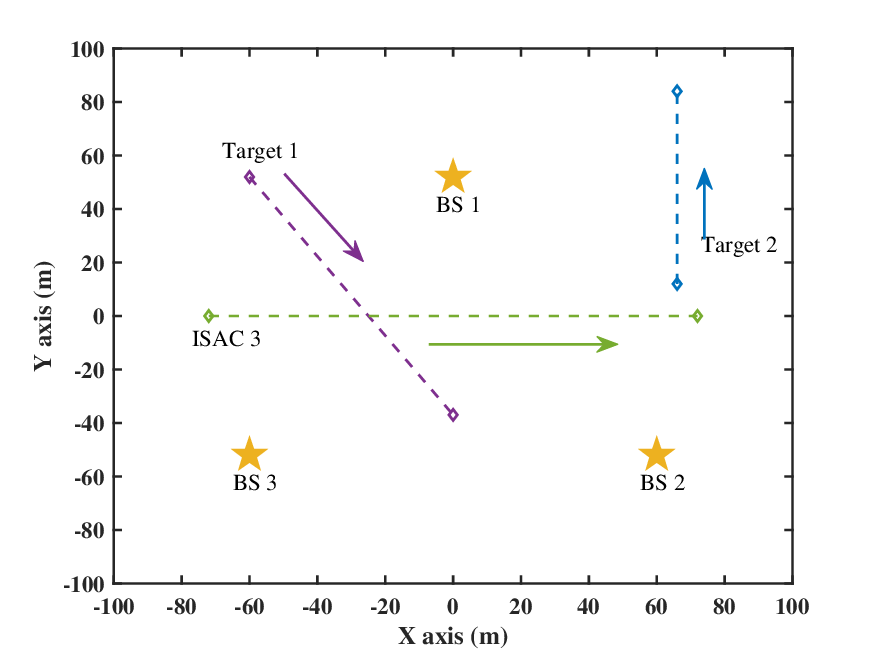}
\caption{Geometric distribution of multiple BSs, sensing targets and ISAC user, where the orrows represent the movement directions of objects and the dashed lines represent the trajectories.}
\label{scen}
\end{figure}

In this section, some simulation results are presented to demonstrate the effectiveness of the proposed CBARA solution. Three benchmark schemes are considered for comparison \cite{yan2018power,xie2017joint}. The geometric distribution of the BSs and the trajectory of targets and users are displayed in Fig. \ref{scen}, while the corresponding initial states are presented in Table \ref{tab2}, where the motion states of the objects can be tracked by particle filtering.
The main simulation parameters are shown in Table \ref{tab3}. Additionally, the transmit power and bandwidth of each object are limited within $\left[0.05\textit{P}_{\mathrm{total}}, 0.85\textit{P}_{\mathrm{total}}\right]$ and $\left[0.05\textit{B}_{\mathrm{total}}, 0.85\textit{B}_{\mathrm{total}}\right]$, respectively. The path-loss for communication users can be obtained through the free-space path loss model $\alpha_{i, k}^{'n} = 32.4+20\mathrm{log}\left(d_{i,k}^n\right)+20\mathrm{log}\left(f_c\right)$.
\begin{table}[h]
\footnotesize
  \begin{center}
  \setlength{\abovecaptionskip}{0.1cm}
    \caption{Initial State}
    \label{tab2}
    \begin{tabular}{c|c|c} 
    \hline
      Object & Initial position (m) & Initial velocity (m/s)  \\
    \hline
        BS 1 &(0.00, 51.96) &/\\
     \hline
         BS 2 &(60.00, -51.96) &/\\
     \hline
         BS 3 &(-60.00, -51.96) &/\\
    \hline
       Target 1& (-60.00, 51.96) & (3.87, -5.73) \\
      \hline
        Target 2& (66.00, 12.00)& (0.00, 4.64)\\
     \hline
         ISAC 3&(-72.00, 9.29)&(9.29, 0.00)\\
      \hline
    \end{tabular}
  \end{center}
\end{table}

\begin{table}[ht!]
\footnotesize
  \begin{center}
  \setlength{\abovecaptionskip}{0.1cm}
    \caption{Simulation Parameters}
    \label{tab3}
    \begin{tabular}{l|r} 
    \hline
      \multicolumn{1}{c|}{Parameters} &  \multicolumn{1}{c}{value}   \\
    \hline 
      Number of BSs $K$ & 3\\
    \hline
      Number of sensing targets $Q$ & 2\\
    \hline
      Number of ISAC users $I$ & 1\\
    \hline
      Number of transmit antennas for each BS $N_t$ & 32\\
    \hline
      Number of receive antennas for each BS $N_r$ & 32\\
    \hline
      Number of receive antennas for each ISAC user $N_r^{'}$ & 2\\
     \hline
     Maximum BS assignment number for each object $L_\mathrm{max}$&3\\
    \hline
    Minimum BS assignment number for each object $L_\mathrm{min}$&2\\
    \hline
      Number of time slots $N$ & 30\\
    \hline
      Sample interval $T_s$ & 0.5 s\\
    \hline
       RCS of each object $|h_{m,k}^n|$ & 1 $\mathrm{m}^2$\\
    \hline
       Carrier frequency $f_c$ & 3 GHz \\
    \hline
       Power spectral density of noise $\sigma_c^2$ &-145 dBm\\
    \hline
       Synchronization error coefficient $\delta_{t}$  & 1.05\\
    \hline
       Total transmit power $\textit{P}_{\mathrm{total}}$ & 30 W \\
    \hline
       Total transmit bandwidth $\textit{B}_{\mathrm{total}}$ & 65 MHz \\
    \hline
       Process noise $\sigma_1$, $\sigma_2$, $\sigma_3$ &1.5, 2, 1 \\
    \hline
       The threshold for BS assignment $\varphi$ & 0.1\\
    \hline
       Number of Monte Carlo trails  &500\\
    \hline
    \end{tabular}
  \end{center}
\end{table}
\subsection{Analysis of resource allocation and BS assignment results }
\begin{figure}[ht!]
\centering
 \setlength{\abovecaptionskip}{0.1cm}
\includegraphics[width=0.5\textwidth]{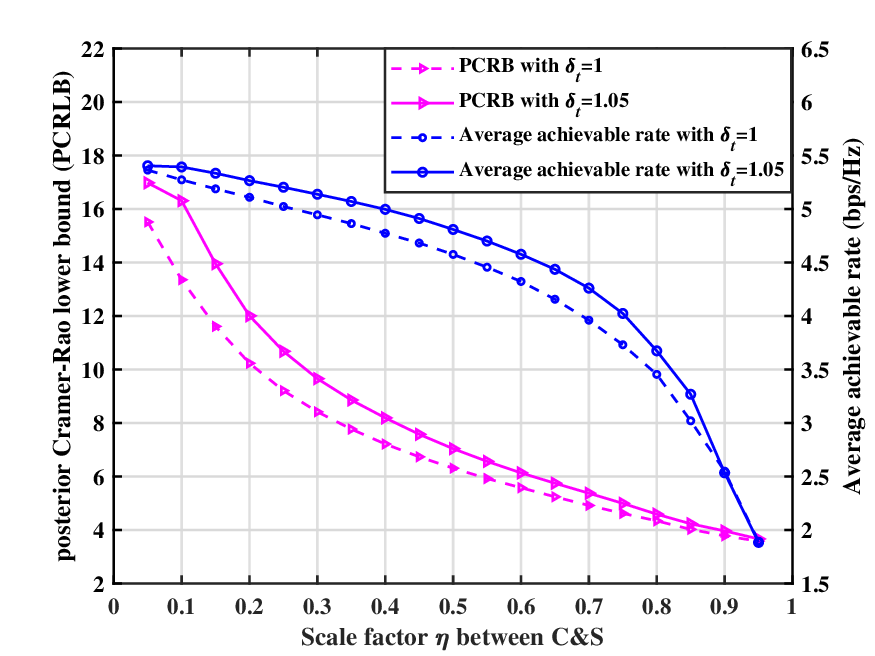}
\caption{C\&S performance with different values of scale factor $\eta$.}
\label{FR_eta}
\end{figure}

Fig. \ref{FR_eta} demonstrates the trade-off between C\&S performance. $\delta_t = 1$ implies a perfect case with no effect of synchronization errors, while $\delta_t = 1.05$ implies that synchronization errors cause a 5\% bias in the accuracy of the parameter estimations. As shown in the figure, the introduction of synchronization errors increases the PCRLB, meaning a decrease in sensing accuracy, but make an improvement in communication performance. As the C\&S scale factor $\eta$ increases from 0.05 to 0.95, the PCRLB drops, the sensing performance improves by about 78\%, yielding higher sensing accuracy. However, in the meantime, the average achievable rate decreases by around 65\%, indicating lower communication performance. 
This trend highlights the effectiveness of the scale factor in optimally adjusting the preference between C\&S according to specific requirements. For example, if balanced C\&S performance is required, such as in smart transportation or urban surveillance systems where both reliable data links and accurate environmental perception are crucial, $\eta$ can be set between 0.4 and 0.7, since the variation of PCRB and average achievable rate are smoother in this interval and both C\&S performances are more biased toward their respective optimal performances. 
$\eta$ greater than 0.7 can be used to favor scenarios where higher sensing performance is required, such as illegal drone detection and autonomous driving, where accurate tracking and localization are essential. 
$\eta$ less than 0.4 is used for scenarios with high data throughput requirements, such as hotspot communication services in densely populated areas, where sensing is provided to assist communication by enabling beam alignment and mobility prediction to enhance communication performance.

\begin{figure}[ht!]
\centering
 \setlength{\abovecaptionskip}{0.1cm}
\includegraphics[width=0.5\textwidth]{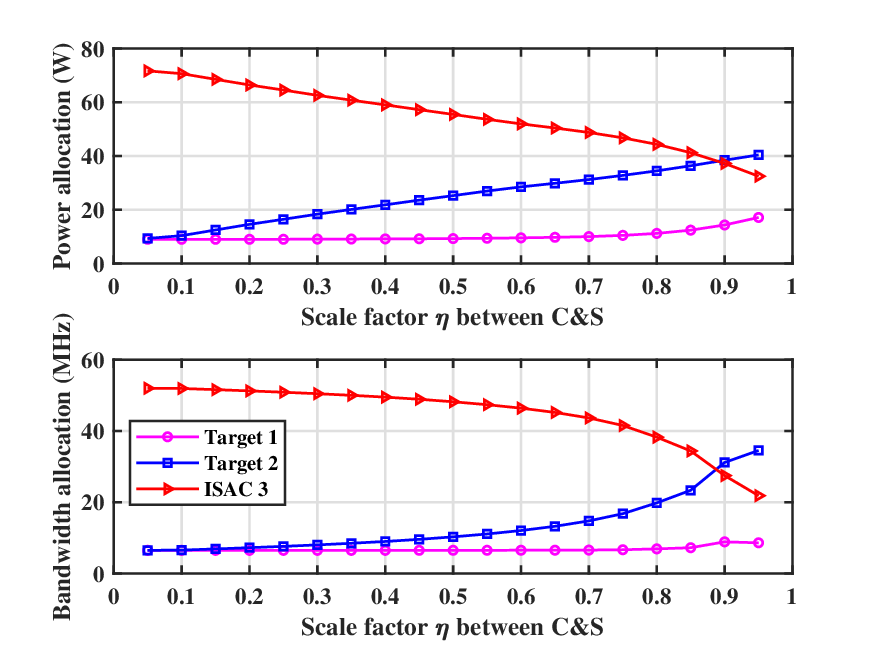}
\caption{Power and bandwidth allocation results with different value of $\eta$ for Target 1, Target 2 and ISAC 3.}
\label{PB_eta}
\end{figure}

The optimized resource allocation results for each scale factor $\eta$ are shown in Fig. \ref{PB_eta}, where the power and bandwidth allocations display a similar trend. In most cases of $\eta$, both resources are prioritized for the ISAC user, since the ISAC user requires more resources to maintain its performance in both C\&S. 
As $\eta$ increases, the power and bandwidth resources allocated to the ISAC user gradually decrease, while those for the sensing targets increase, because a larger $\eta$ corresponds to a lower requirement for communication performance and a higher demand for sensing performance.
Additionally, the power and bandwidth allocated to Target 2 increases significantly faster than those to Target 1, because Target 2's trajectory moves away from the network center, resulting in decreasing sensing channel gain. Consequently, more resources are required to satisfy the sensing performance demand. 
\begin{figure}[ht!]
\centering
 \setlength{\abovecaptionskip}{0.1cm}
\includegraphics[width=0.5\textwidth]{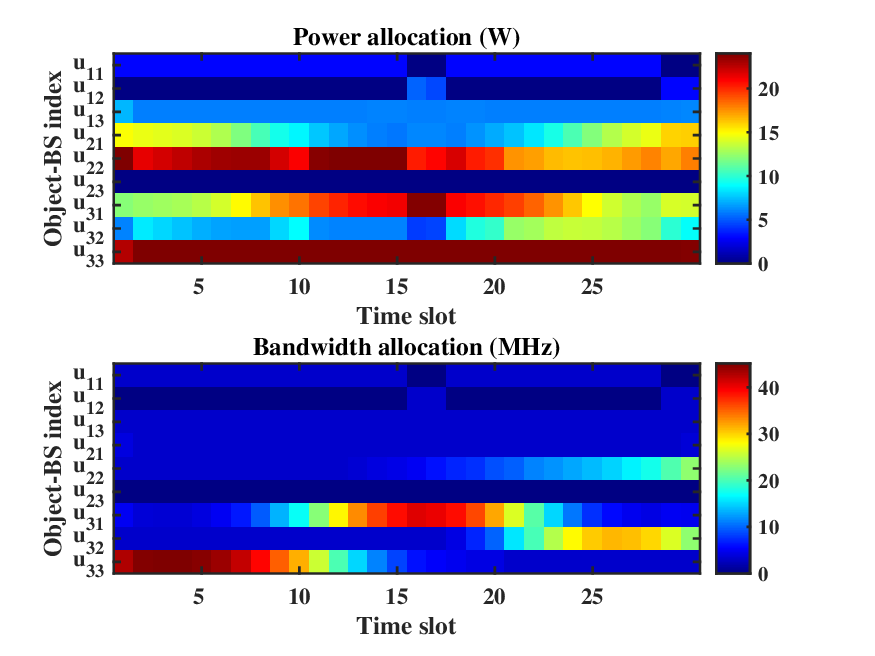}
\caption{BS assignment and resource allocation results with $\eta$ = 0.7.}
\label{pandb_allocation0.7}
\end{figure}
\begin{figure}[ht!]
\centering
 \setlength{\abovecaptionskip}{0.1cm}
\includegraphics[width=0.5\textwidth]{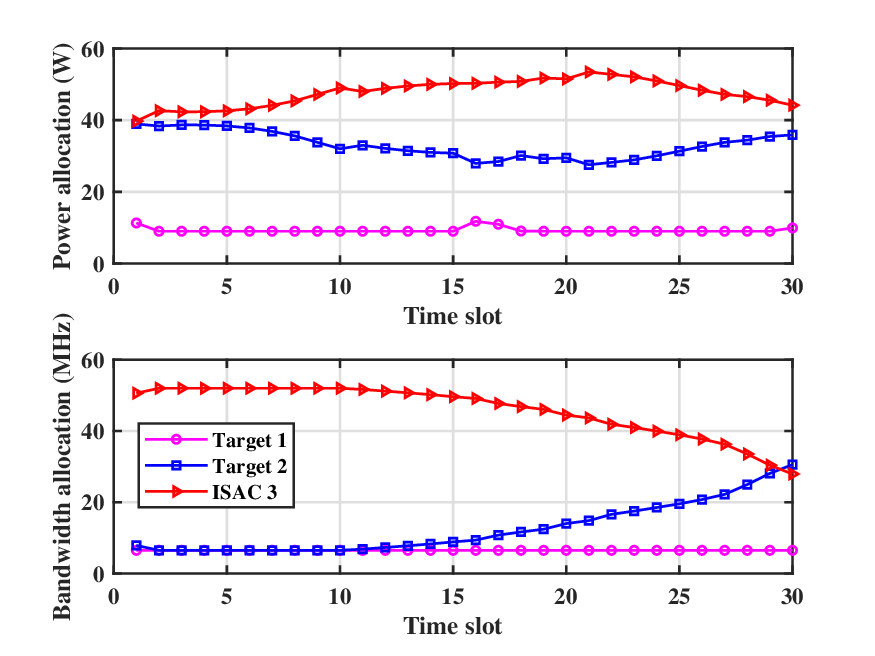}
\caption{Power and bandwidth allocation results for sensing targets and ISAC user with $\eta$ = 0.7.}
\label{competition}
\end{figure}

Fig. \ref{pandb_allocation0.7} shows the optimization results of CBARA during the mission period. To avoid excessive imbalance between C\&S performance due to too large or too small $\eta$, and to make the results of resource allocation easier to analyze, we choose the case of $\eta = 0.7$ for analysis. In Fig. \ref{pandb_allocation0.7}, the vertical scale $\mathrm{u}_{mk}$ denotes the allocation relationship between the $m$-th object and the $k$-th BS. Within each time slot, the deep blue region signifies $u_{m, k}=0$, while regions in other colors denote $u_{m, k}=1$. The colors of the rectangles represent the values of allocated power and bandwidth according to the respective colorbars. Analyzing the BSs allocation outcomes, it is observed that throughout the mission duration, all BSs are assigned to the ISAC user to ensure its stable communication performance. Moreover, BS 1 and BS 2 are consistently assigned to Target 2, while BS 1 and BS 3 are predominantly assigned to Target 1 for the majority of the period. This phenomenon is largely caused by the distance from each of the two targets to the three BSs, respectively. 
In addition, the trends of power and bandwidth allocations are inconsistent throughout the mission time slot because power and bandwidth have different effects on PCRLB and achievable rate, which can be derived from the derivation in Section II.
Furthermore, there is noticeable resource competition between ISAC 3 and Target 2, as intuitively observed in Fig. \ref{competition}. This is attributed to their simultaneous sharing of resources from BS 1 and BS 2. Specifically, before the 15-th time slot, the bandwidth resources for ISAC 3 are provided mainly by BS 1 and BS 3 as shown in Fig. \ref{pandb_allocation0.7}. However, after the 15-th time slot, the bandwidth resources for ISAC 3 are mainly from BS 1 and BS 2, intensifying the competition between Target 2 and ISAC 3. As Target 2 moves away from the network center in Fig. \ref{scen}, its channel conditions deteriorate, requiring more bandwidth resources to maintain its sensing performance. Consequently, the bandwidth resources allocated to Target 2 gradually increase after the 15-th time slot and eventually exceed those allocated to ISAC 3. 
\begin{figure}[ht!]
\centering
 \setlength{\abovecaptionskip}{0.1cm}
\includegraphics[width=0.5\textwidth]{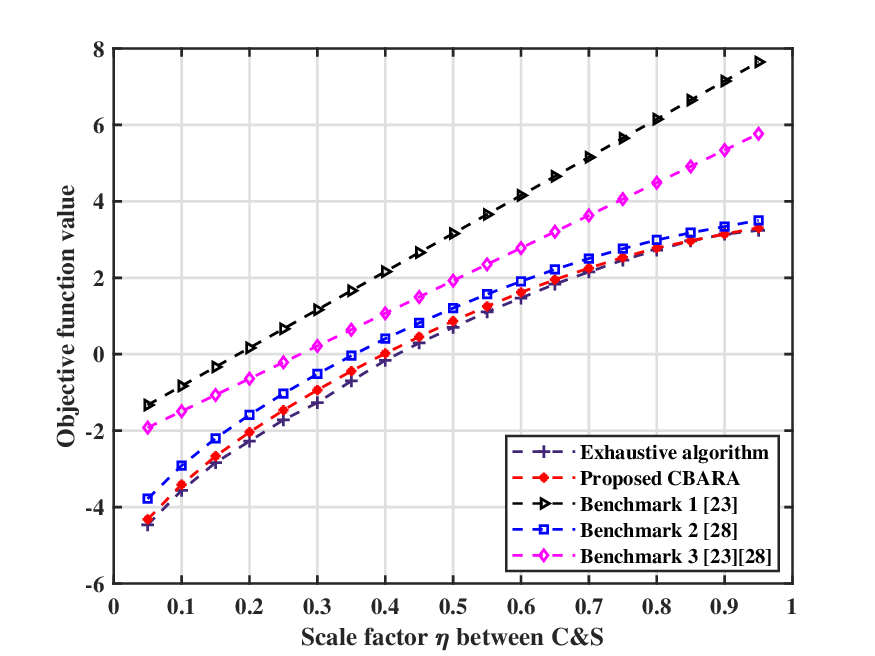}
\caption{Objective function value for five schemes with different value of scale factor $\eta$ when $\delta_t=1$.}
\label{eta_objective_value}
\end{figure}

\subsection{Effectiveness of the CBARA Strategy}
To evaluate the performance of our proposed CBARA strategy, we have conceived the following three schemes as benchmarks.
\begin{itemize}
\item{\textit{Exhaustive algorithm}: In this scheme, all possible BS assignment combinations for each object are enumerated based on the BS assignment constraints in \eqref{P1}. Then, for each combination, resource allocation is optimized by AO method. Finally, the combination with the minimum objective function value among all combinations is selected as the globally optimal solution. This algorithm serves as a performance upper bound for evaluating the effectiveness of the proposed method.}
\item{\textit{Benchmark 1 \cite{yan2018power}}: In this scheme, all BSs are assigned to each object (i.e., $u_{m,k}=1$ for all $m,k$), and both the transmit power and bandwidth are uniformly allocated among objects without optimization, which represents full connectivity and equal resource sharing, serving as a benchmark for other optimization schemes.}
\item{\textit{Benchmark 2 \cite{xie2017joint}}: This scheme also assumes that all BSs serve all objects (i.e., $u_{m,k}=1$ for all $m,k$), but the power and bandwidth allocation is jointly optimized using the same AO method as in the proposed CBARA strategy. This allows us to isolate the impact of BS assignment in order to analyze the performance improvement brought solely by resource optimization.}
\item{\textit{Benchmark 3 \cite{yan2018power,xie2017joint}}: This scheme employs the proposed heuristic algorithm for BS assignment with uniform power and bandwidth allocation across all objects. The effect of BS assignment can be evaluated independently of resource optimization.}
\end{itemize}
For all four benchmark schemes, the remaining system parameters are identical to those listed in Table \ref{tab3}, unless explicitly redefined, ensuring consistency across all simulations.

Fig. \ref{eta_objective_value} illustrates the objective function value versus the trade-off scale factor $\eta$ under five different schemes, where the objective function value embodies the ability to trade-off C\&S and the efficiency of resource utilization. In this figure, all curves show an increasing trend with respect to $\eta$, since the sensing term dominates the objective when $\eta$ increases, while the communication term decreases, contributing to a higher value, which aligns with the formulation in \eqref{objective_function}.
When comparing the different algorithms, firstly, benchmark 3, an optimized approach for BS assignment only, presents an improved objective function value over benchmark 1, an approach that uniformly allocates both BSs and resources, indicating that our optimal BS assignment strategy can significantly enhance the resource utilization efficiency of the system. This conclusion can also be observed by comparing the proposed CBARA with benchmark 2, an approach that only optimizes resource allocation. Secondly, it is evident that the performance improvement achieved through resource allocation is more significant than that achieved through BS assignment, as seen when comparing the gain between benchmark 1 and benchmark 2 with those between benchmark 1 and benchmark 3. This phenomenon can be attributed to the fact that power and bandwidth allocation provide greater flexibility for optimization in adjusting the trade-off between communication and sensing performance. For this reason, the proposed joint optimization of BS assignment and resource allocation achieves superior performance compared to the other three schemes. Finally, the performance of CBARA closely approaches that of the exhaustive search algorithm, which represents the near-optimal solution but suffers from significantly higher computational complexity, suggesting that our proposed CBARA offers a favorable trade-off between performance and computational efficiency.

\begin{figure}[ht!]
\centering
 \setlength{\abovecaptionskip}{0.1cm}
\includegraphics[width=0.5\textwidth]{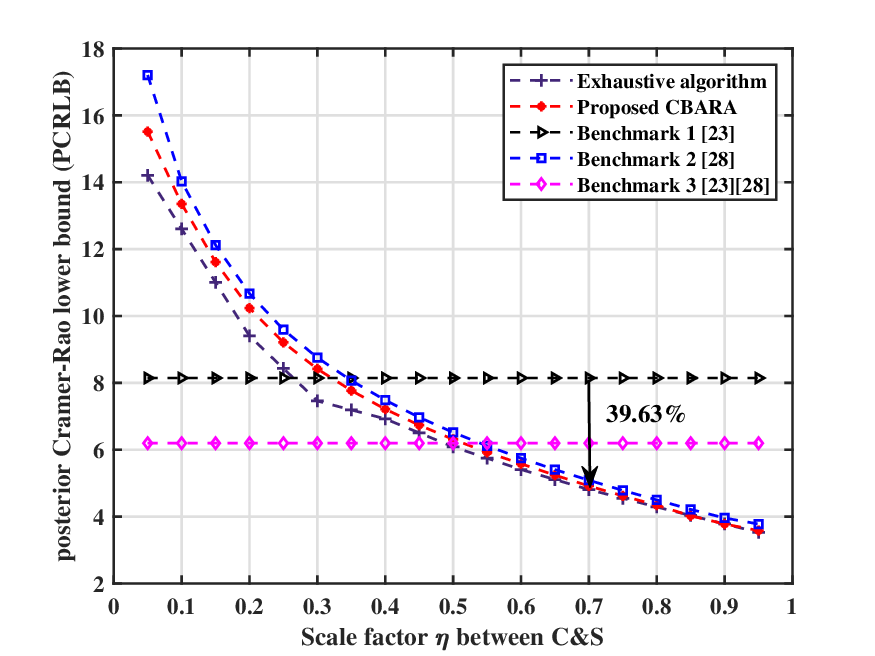}
\caption{PCRLB for five schemes with different value of scale factor $\eta$ when $\delta_t=1$.}
\label{pcrbcompare}
\end{figure}
\begin{figure}[ht!]
\centering
 \setlength{\abovecaptionskip}{0.1cm}
\includegraphics[width=0.5\textwidth]{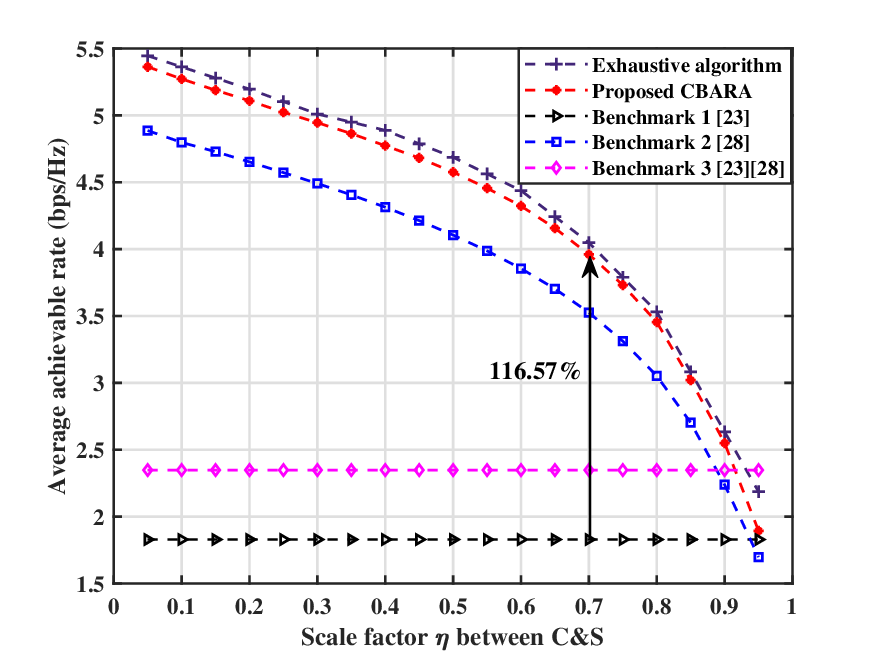}
\caption{Average achievable rate for five schemes with different value of scale factor $\eta$ when $\delta_t=1$.}
\label{ratecompare}
\end{figure}
Fig. \ref{pcrbcompare} and Fig. \ref{ratecompare} display the comparative results of C\&S performance across the five different schemes. As shown in Fig. \ref{pcrbcompare}, the gain in sensing performance due to BS assignment follows the same trend as in Fig. \ref{eta_objective_value}. For comparison of resource allocation schemes, in cases with small $\eta$, the sensing performance of the proposed scheme, benchmark 2, and exhaustive algorithm is inferior to that of benchmark 3 and benchmark 1, respectively, suggesting that the resources allocated to the sensing targets are less than the average resources with a bias in favor of communication performance. However, in cases with large $\eta$, our proposed scheme exhibits superior sensing performance, closely approaching the exhaustive algorithm.

In Fig. \ref{ratecompare}, for the BS assignment schemes, benchmark 3 also demonstrates better communication performance than benchmark 1, which, together with Fig. \ref{eta_objective_value} and Fig. \ref{pcrbcompare}.
For resource allocation schemes, the communication performance of our proposed scheme and benchmark 2 consistently outperforms that of benchmark 3 and benchmark 1, respectively, when $\eta$ is below 0.9. From the above comparisons, it can be concluded that even if the optimized resource allocation scheme does not provide the best performance in both communication and sensing in all cases of $\eta$, it has a greater advantage when comprehensively assessing both C\&S performance. Furthermore, it is noted that the BS assignment schemes as benchmark 1 and benchmark 3 are insensitive to $\eta$, as their C\&S performance remains almost stable with the increase of $\eta$. However, in the proposed scheme and the benchmark 2, $\eta$ directly affects PCRLB and achievable rate which are determined by the results of power and bandwidth allocation, leading to significant variations. In general, our proposed CBARA scheme exhibits superiority in balancing C\&S performance, especially when compared to benchmark 1, with performance improvements of about 117\%  and 40\% in communication and sensing, respectively, at $\eta=0.7$. 
\begin{figure} [h]
\centering
 \setlength{\abovecaptionskip}{0.1cm}
\includegraphics[width=0.5\textwidth]{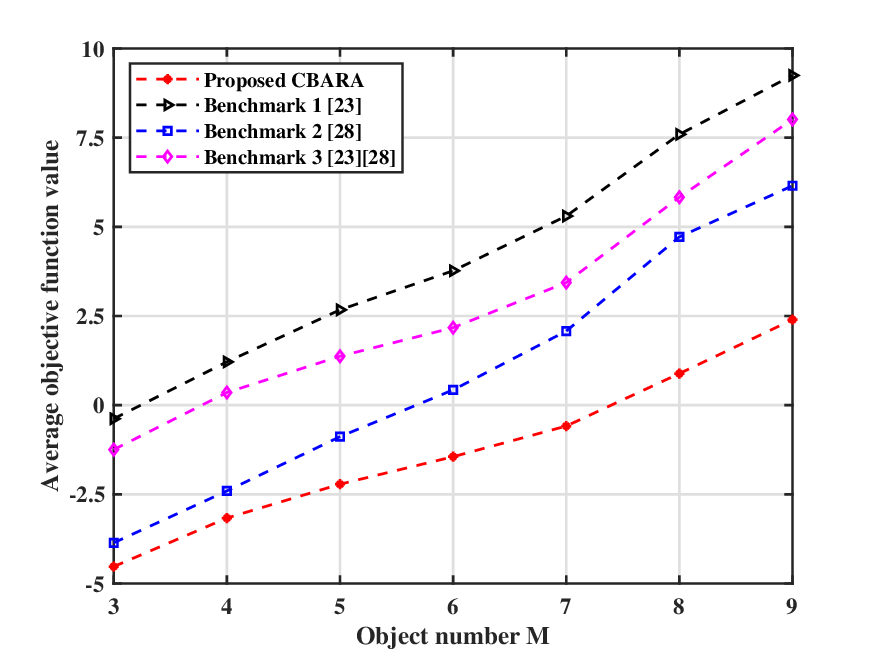}
\caption{Average objective function value for four schemes with different object number $M$ when $P_{\mathrm{total}}=90$ W and $B_{\mathrm{total}}=120$ MHz.}
\label{object_num_value}
\end{figure}

Fig. \ref{object_num_value} presents the average objective function value for four different schemes across varying numbers of objects $M$, with $P_{\mathrm{total}}=90$ W and $B_{\mathrm{total}}=120$ MHz. As the number of service objects increases, the objective function value also rises, since the demand for resources grows while the total resource capacity remains fixed, which leads to a decline in overall sensing and communication performance, with an increase in PCRLB and a decrease in achievable rate. Among the four schemes, the proposed CBARA strategy consistently maintains the lowest objective function value, and the performance comparison between the other three schemes aligns with the results shown in Fig. \ref{eta_objective_value}. Furthermore, the gap between our proposed CBARA and other schemes widens further as the system scale grows, highlighting the superiority of our strategy in handling larger-scale systems and its effectiveness in enhancing comprehensive C\&S performance.

\section{Conclusion}
In this paper, we consider a multi-base station ISAC network that enables communication while achieving multi-target sensing. We therefore propose a CBARA strategy to jointly optimize C\&S performance. Initially, the achievable rate and the PCRLB for multi-BS multi-target sensing, containing power and bandwidth, are derived as the communication and sensing performance indicators, respectively. Subsequently, the proposed CBARA scheme is formulated as an optimization problem, revealing that the issue is a non-convex problem with three variables. Consequently, a heuristic alternating optimization algorithm is introduced to obtain a sub-optimal solution for the optimization problem. Finally, simulation results validate the effectiveness and superiority of our proposed approach. 

Future work might concern more complex scenarios, such as imperfect beamforming and interference control in the network. Moreover, approaches involving deep learning and reinforcement learning will be considered for incorporation into the architecture of intelligent networks.
\section*{Appendix}
In this section, we prove the convexity of the objective function \eqref{P1.1} with respect to the variables ${\bf{p}}$ or ${\bf{b}}$ for given ${\bf{u}}$.

Since the linear combination of convex functions is still convex \cite{convex}, the proof is divided into the following two parts
\begin{equation}\label{C}
    C\left({\bf{p}}_m, {\bf{b}}_m\right)=\mathrm{tr}\left( {{\bf{J}}_{{\bf{x}}_m}^{ - 1}\left( {{\bf{p}}_m,{\bf{b}}_m} \right)} \right),
\end{equation}
and
\begin{equation}\label{G}
    G\left(p_{i, k}, b_{i, k}\right)=-b_{i, k} \log _2\left(1+\frac{p_{i, k} h_{i, k}}{b_{i, k}}\right),
\end{equation}
where $h_{i,k}=\varsigma_{i, k}/\sigma_c^2$.

Firstly, for a fixed ${\bf{b}}_m$, \eqref{C} can be rewritten as 
\begin{align}\label{C1}
    C\left({\bf{p}}_m\right)&=\mathrm{tr}\left\{\left[\mathbf{J}_P+\sum_{k \in \mathcal{S}_{\mathrm{o},m}}\mathbf{H}_{m, k}^T\mathbf{\Lambda}_{m, k}^{-1}\left(p_{m, k}\right) \mathbf{H}_{m, k} \right]^{-1}\right\}\nonumber\\  
   &=\mathrm{tr}\left\{\left[\mathbf{J}_P+\sum_{k \in \mathcal{S}_{\mathrm{o},m}}\sum_{j=1}^3 p_{m,k}a_j{\bf{h}}_j{\bf{h}}_j^T \right]^{-1}\right\},
\end{align}
where $a_j$ denotes the $j$-th diagonal element of the remaining part of $\mathbf{\Lambda}_{m, k}^{-1}$ after extracting $p_{m, k}$, ${\bf{h}}_j$ is the $j$-th column of $\mathbf{H}_{m, k}$.
For any two chosen variables ${\bf{p}}_m^1$, ${\bf{p}}_m^2$, and an arbitrary $\alpha\in\left[0,1\right]$, we have
 \begin{align}\label{C2} 
        &C\left(\alpha \mathbf{p}_m^1 + (1-\alpha) \mathbf{p}_m^2\right) \nonumber\\
        &= \mathrm{tr}\left\{ \left[ \mathbf{J}_P + \sum_{k \in \mathcal{S}_{\mathrm{o},m}} \sum_{j=1}^3 \left( \alpha p_{m,k}^1 + (1-\alpha) p_{m,k}^2 \right) a_j \mathbf{h}_j \mathbf{h}_j^T \right]^{-1} \right\} \nonumber\\
        &= \mathrm{tr}\left\{ \left[ \alpha \mathbf{J}_P + \sum_{k \in \mathcal{S}_{\mathrm{o},m}} \sum_{j=1}^3 \alpha p_{m,k}^1 a_j \mathbf{h}_j \mathbf{h}_j^T \right. \right. \nonumber\\
        &\quad \left. \left. + (1-\alpha) \mathbf{J}_P + \sum_{k \in \mathcal{S}_{\mathrm{o},m}} \sum_{j=1}^3 (1-\alpha) p_{m,k}^2 a_j \mathbf{h}_j \mathbf{h}_j^T \right]^{-1} \right\}\nonumber\\
        &\le \alpha C\left(\mathbf{p}_m^1\right)+\left(1-\alpha\right)C\left(\mathbf{p}_m^2\right).
\end{align}
Since ${\bf{J}}_{{\bf{x}}_m}$ is a positive semidefinite matrix, $\mathrm{tr}\left({\bf{J}}_{{\bf{x}}_m}^{-1}\right)$ is a convex function and the inequality in \eqref{C2} holds true \cite{convex}. Subsequently, as any arbitrary $\mathbf{p}_m$ is positive, $C\left({\bf{p}}_m\right)$ is a convex function with respect to $\mathbf{p}_m$.

Similarly, for a fixed $\mathbf{p}_m$, \eqref{C} can be reformulated as
\begin{align}\label{C3}
    &C\left({\bf{b}}_m\right)\\
    &=\mathrm{tr}\left\{\left[\mathbf{J}_P+\sum_{k \in \mathcal{S}_{\mathrm{o},m}}\mathbf{H}_{m, k}^T\mathbf{\Lambda}_{m, k}^{-1}\left(b_{m, k}\right) \mathbf{H}_{m, k} \right]^{-1}\right\}\nonumber\\  
   &=\mathrm{tr}\left\{\left[\mathbf{J}_P+\sum_{k \in \mathcal{S}_{\mathrm{o},m}}\left(b_{m,k}a_1^{'}{\bf{h}}_1{\bf{h}}_1^T 
   +\sum_{j=2}^3 a_j^{'}{\bf{h}}_j{\bf{h}}_j^T\right) \right]^{-1}\right\},
\end{align}
where $a_j^{'}$ is the $j$-th diagonal element of $\mathbf{\Lambda}_{m, k}^{-1}$ after extraction. Obviously, the convexity of \eqref{C3} can be proven by a process similar to \eqref{C2}.

Secondly, we prove that \eqref{G} is a convex function for $p_{i,k}$ or $b_{i,k}$. The second order derivatives of the function $G\left(p_{i,k}, b_{i,k}\right)$ with respect to $p_{i,k}$ and $b_{i,k}$ are calculated as 
\begin{align}
\frac{\partial^2 G}{\partial p_{i,k}^2}=\frac{h_{i,k}^2}{b_{i,k}\left(1+\frac{h_{i,k}p_{i,k}}{b_{i,k}}\right)^2 \ln2},\\
\frac{\partial^2 G}{\partial b_{i,k}^2}=\frac{h_{i,k}^2p_{i,k}^2}{b_{i,k}^3\left(1+\frac{h_{i,k}p_{i,k}}{b_{i,k}}\right)^2 \ln2}.
\end{align}
Since $p_{i,k}$, $b_{i,k}$ and $h_{i,k}$ are all positive valued, $G\left(p_{i,k}\right)$ and $G\left(b_{i,k}\right)$ are convex. The objective function \eqref{P1.1} is expressed as a linear combination of \eqref{C} and \eqref{G}, hence it is also convex.

\bibliographystyle{IEEEtran}
\bibliography{reference}

\begin{thebibliography}{10}
\providecommand{\url}[1]{#1}
\csname url@samestyle\endcsname
\providecommand{\newblock}{\relax}
\providecommand{\bibinfo}[2]{#2}
\providecommand{\BIBentrySTDinterwordspacing}{\spaceskip=0pt\relax}
\providecommand{\BIBentryALTinterwordstretchfactor}{4}
\providecommand{\BIBentryALTinterwordspacing}{\spaceskip=\fontdimen2\font plus
\BIBentryALTinterwordstretchfactor\fontdimen3\font minus \fontdimen4\font\relax}
\providecommand{\BIBforeignlanguage}[2]{{%
\expandafter\ifx\csname l@#1\endcsname\relax
\typeout{** WARNING: IEEEtran.bst: No hyphenation pattern has been}%
\typeout{** loaded for the language `#1'. Using the pattern for}%
\typeout{** the default language instead.}%
\else
\language=\csname l@#1\endcsname
\fi
#2}}
\providecommand{\BIBdecl}{\relax}
\BIBdecl

\bibitem{lu2024integrated}
S.~Lu, F.~Liu, Y.~Li, K.~Zhang, H.~Huang, J.~Zou, X.~Li, Y.~Dong, F.~Dong, J.~Zhu \emph{et~al.}, ``Integrated sensing and communications: Recent advances and ten open challenges,'' \emph{IEEE Internet Things J.}, vol.~11, no.~11, pp. 19\,094--19\,120, Jun. 2024.

\bibitem{kaushik2024toward}
A.~Kaushik, R.~Singh, S.~Dayarathna, R.~Senanayake, M.~Di~Renzo, M.~Dajer, H.~Ji, Y.~Kim, V.~Sciancalepore, A.~Zappone \emph{et~al.}, ``Toward integrated sensing and communications for {6G}: Key enabling technologies, standardization, and challenges,'' \emph{IEEE Commun. Standards Mag.}, vol.~8, no.~2, pp. 52--59, Jun. 2024.

\bibitem{ISAC}
F.~Liu, Y.~Cui, C.~Masouros, J.~Xu, T.~X. Han, Y.~C. Eldar, and S.~Buzzi, ``\BIBforeignlanguage{en-US}{Integrated sensing and communications: Toward dual-functional wireless networks for {6G} and beyond},'' \emph{\BIBforeignlanguage{en-US}{IEEE J. Sel. Areas Commun.}}, p. 1728–1767, Jun. 2022.

\bibitem{xiang2023green}
L.~Xiang, K.~Xu, J.~Hu, and K.~Yang, ``Green beamforming design for integrated sensing and communication systems: A practical approach using beam-matching error metrics,'' \emph{IEEE Trans. Veh. Technol.}, vol.~73, no.~4, pp. 5935--5940, Apr. 2023.

\bibitem{Applications2020joint}
F.~Liu, C.~Masouros, A.~P. Petropulu, H.~Griffiths, and L.~Hanzo, ``Joint radar and communication design: Applications, state-of-the-art, and the road ahead,'' \emph{IEEE Trans. Commun.}, vol.~68, no.~6, pp. 3834--3862, Jun. 2020.

\bibitem{autonomous2020joint}
D.~Ma, N.~Shlezinger, T.~Huang, Y.~Liu, and Y.~C. Eldar, ``Joint radar-communication strategies for autonomous vehicles: Combining two key automotive technologies,'' \emph{IEEE Signal Process. Mag.}, vol.~37, no.~4, pp. 85--97, Jul. 2020.

\bibitem{liu2022communication}
Y.~Liu and K.~Yang, ``Communication, sensing, computing and energy harvesting in smart cities,'' \emph{IET Smart Cities}, vol.~4, no.~4, pp. 265--274, Sep. 2022.

\bibitem{DongSensing2024}
F.~Dong, F.~Liu, Y.~Cui, S.~Lu, and Y.~Li, ``Sensing as a service in {6G} perceptive mobile networks: Architecture, advances, and the road ahead,'' \emph{IEEE Netw.}, vol.~38, no.~2, pp. 87--96, Mar. 2024.

\bibitem{kivanc2003computationally}
D.~Kivanc, G.~Li, and H.~Liu, ``Computationally efficient bandwidth allocation and power control for {OFDMA},'' \emph{IEEE Trans. Wirel. Commun.}, vol.~2, no.~6, pp. 1150--1158, Nov. 2003.

\bibitem{tachwali2013multiuser}
Y.~Tachwali, B.~F. Lo, I.~F. Akyildiz, and R.~Agusti, ``Multiuser resource allocation optimization using bandwidth-power product in cognitive radio networks,'' \emph{IEEE J. Sel. Areas Commun.}, vol.~31, no.~3, pp. 451--463, Mar. 2013.

\bibitem{zhang2017downlink}
H.~Zhang, H.~Liu, J.~Cheng, and V.~C. Leung, ``Downlink energy efficiency of power allocation and wireless backhaul bandwidth allocation in heterogeneous small cell networks,'' \emph{IEEE Trans. Commun.}, vol.~66, no.~4, pp. 1705--1716, Apr. 2017.

\bibitem{jiang2019joint}
Y.~Jiang, Y.~Zou, H.~Guo, T.~A. Tsiftsis, M.~R. Bhatnagar, R.~C. de~Lamare, and Y.-D. Yao, ``Joint power and bandwidth allocation for energy-efficient heterogeneous cellular networks,'' \emph{IEEE Trans. Commun.}, vol.~67, no.~9, pp. 6168--6178, Sep. 2019.

\bibitem{MIMORadarLocalization}
N.~Garcia, A.~M. Haimovich, M.~Coulon, and M.~Lops, ``Resource allocation in {MIMO} radar with multiple targets for non-coherent localization,'' \emph{IEEE Trans. Signal Process.}, vol.~62, no.~10, pp. 2656--2666, May 2014.

\bibitem{yan2015simultaneous}
J.~Yan, H.~Liu, B.~Jiu, B.~Chen, Z.~Liu, and Z.~Bao, ``Simultaneous multibeam resource allocation scheme for multiple target tracking,'' \emph{IEEE Trans. Signal Process.}, vol.~63, no.~12, pp. 3110--3122, Jun. 2015.

\bibitem{zhang2020power}
H.~Zhang, B.~Zong, and J.~Xie, ``Power and bandwidth allocation for multi-target tracking in collocated {MIMO} radar,'' \emph{IEEE Trans. Veh. Technol.}, vol.~69, no.~9, pp. 9795--9806, Sep. 2020.

\bibitem{Luong2021RadioResource}
N.~C. Luong, X.~Lu, D.~T. Hoang, D.~Niyato, and D.~I. Kim, ``Radio resource management in joint radar and communication: A comprehensive survey,'' \emph{IEEE Commun. Surv. Tutor.}, vol.~23, no.~2, pp. 780--814, 2nd Quart. 2021.

\bibitem{zhou2018resource}
Y.~Zhou, H.~Zhou, F.~Zhou, Y.~Wu, and V.~C. Leung, ``Resource allocation for a wireless powered integrated radar and communication system,'' \emph{IEEE Wirel. Commun. Lett.}, vol.~8, no.~1, pp. 253--256, Feb. 2018.

\bibitem{ahmed2024distributed}
A.~Ahmed and Y.~D. Zhang, ``Optimized resource allocation for distributed joint radar-communication system,'' \emph{IEEE Trans. Veh. Technol.}, vol.~73, no.~3, pp. 3872--3885, Mar. 2024.

\bibitem{xiang2024robust}
L.~Xiang, K.~Xu, J.~Hu, C.~Masouros, and K.~Yang, ``Robust {NOMA}-assisted {OTFS-ISAC} network design with {3D} motion prediction topology,'' \emph{IEEE Internet Things J.}, vol.~11, no.~9, pp. 15\,909--15\,918, May 2024.

\bibitem{9729765}
C.~Ding, J.-B. Wang, H.~Zhang, M.~Lin, and G.~Y. Li, ``Joint {MIMO} precoding and computation resource allocation for dual-function radar and communication systems with mobile edge computing,'' \emph{IEEE J. Sel. Areas Commun.}, vol.~40, no.~7, pp. 2085--2102, Jul. 2022.

\bibitem{dong2022sensing}
F.~Dong, F.~Liu, Y.~Cui, W.~Wang, K.~Han, and Z.~Wang, ``Sensing as a service in {6G} perceptive networks: A unified framework for {ISAC} resource allocation,'' \emph{IEEE Trans. Wirel. Commun.}, vol.~22, no.~5, pp. 3522--3536, May 2023.

\bibitem{li2024maximizing}
B.~Li, X.~Wang, and F.~Fang, ``Maximizing the value of service provisioning in multi-user {ISAC} systems through fairness guaranteed collaborative resource allocation,'' \emph{IEEE J. Sel. Areas Commun.}, vol.~42, no.~9, pp. 2243--2258, Sep. 2024.

\bibitem{yan2018power}
J.~Yan, H.~Liu, and Z.~Bao, ``Power allocation scheme for target tracking in clutter with multiple radar system,'' \emph{Signal Process.}, vol. 144, pp. 453--458, Mar. 2018.

\bibitem{wei2023integrated}
Z.~Wei, W.~Jiang, Z.~Feng, H.~Wu, N.~Zhang, K.~Han, R.~Xu, and P.~Zhang, ``Integrated sensing and communication enabled multiple base stations cooperative sensing towards {6G},'' \emph{IEEE Netw.}, Oct. 2023.

\bibitem{meng2024cooperative}
K.~Meng, C.~Masouros, A.~P. Petropulu, and L.~Hanzo, ``Cooperative {ISAC} networks: Performance analysis, scaling laws and optimization,'' \emph{arXiv preprint arXiv:2404.14514}, Jun. 2024.

\bibitem{wei2024deepCooperation}
Z.~Wei, H.~Liu, Z.~Feng, H.~Wu, F.~Liu, and Q.~Zhang, ``Deep cooperation in {ISAC} system: Resource, node and infrastructure perspectives,'' \emph{IEEE Internet Things Mag.}, vol.~7, no.~6, pp. 118--125, Nov. 2024.

\bibitem{tharmarasa2011decentralized}
R.~Tharmarasa, T.~Kirubarajan, A.~Sinha, and T.~Lang, ``Decentralized sensor selection for large-scale multisensor-multitarget tracking,'' \emph{IEEE Trans. Aerosp. Electron. Syst.}, vol.~47, no.~2, pp. 1307--1324, Apr. 2011.

\bibitem{xie2017joint}
M.~Xie, W.~Yi, T.~Kirubarajan, and L.~Kong, ``Joint node selection and power allocation strategy for multitarget tracking in decentralized radar networks,'' \emph{IEEE Trans. Signal Process.}, vol.~66, no.~3, pp. 729--743, Feb. 2018.

\bibitem{yi2020resource}
W.~Yi, Y.~Yuan, R.~Hoseinnezhad, and L.~Kong, ``Resource scheduling for distributed multi-target tracking in netted colocated {MIMO} radar systems,'' \emph{IEEE Trans. Signal Process.}, vol.~68, pp. 1602--1617, Feb. 2020.

\bibitem{chen2015cooperative}
H.~Chen, S.~Ta, and B.~Sun, ``Cooperative game approach to power allocation for target tracking in distributed {MIMO} radar sensor networks,'' \emph{IEEE Sens. J.}, vol.~15, no.~10, pp. 5423--5432, Oct. 2015.

\bibitem{meng2024cooperativeISAC}
K.~Meng, C.~Masouros, A.~P. Petropulu, and L.~Hanzo, ``Cooperative {ISAC} networks: Opportunities and challenges,'' \emph{IEEE Wirel. Commun.}, vol.~32, no.~3, pp. 212--219, Jun. 2025.

\bibitem{XiaSymbiotic}
F.~Xia, Z.~Fei, X.~Wang, W.~Yuan, Q.~Wu, Y.~Liu, and T.~Q.~S. Quek, ``Symbiotic sensing and communication: Framework and beamforming design,'' \emph{IEEE Trans. Wirel. Commun.}, vol.~24, no.~3, pp. 2417--2434, Mar. 2025.

\bibitem{Meng2023Performance}
M.~Liu, M.~Yang, H.~Li, K.~Zeng, Z.~Zhang, A.~Nallanathan, G.~Wang, and L.~Hanzo, ``Performance analysis and power allocation for cooperative {ISAC} networks,'' \emph{IEEE Internet Things J.}, vol.~10, no.~7, pp. 6336--6351, Apr. 2023.

\bibitem{Coordinated2022}
Y.~Huang, Y.~Fang, X.~Li, and J.~Xu, ``Coordinated power control for network integrated sensing and communication,'' \emph{IEEE Trans. Veh. Technol.}, vol.~71, no.~12, pp. 13\,361--13\,365, Dec. 2022.

\bibitem{zhang2023joint}
J.~Zhang, Z.~Fei, X.~Wang, P.~Liu, J.~Huang, and Z.~Zheng, ``Joint resource allocation and user association for multi-cell integrated sensing and communication systems,'' \emph{EURASIP J. Wirel. Commun. Netw.}, vol. 2023, no.~1, p.~64, Jul. 2023.

\bibitem{LiuOffloading}
P.~Liu, Z.~Fei, X.~Wang, Y.~Zhou, Y.~Zhang, and F.~Liu, ``Joint beamforming and offloading design for integrated sensing, communication, and computation system,'' \emph{IEEE Trans. Veh. Technol.}, pp. 1--6, Apr. 2025.

\bibitem{JiaxingAd-Hoc}
J.~Wang, L.~Bai, J.~Chen, and J.~Wang, ``Starling flocks-inspired resource allocation for {ISAC}-aided green ad hoc networks,'' \emph{IEEE Trans. Green Commun. Netw.}, vol.~7, no.~1, pp. 444--454, Mar. 2023.

\bibitem{meng2022multi}
K.~Meng, X.~He, Q.~Wu, and D.~Li, ``Multi-{UAV} collaborative sensing and communication: Joint task allocation and power optimization,'' \emph{IEEE Trans. Wirel. Commun.}, vol.~22, no.~6, pp. 4232--4246, Jun. 2022.

\bibitem{pan2024cooperativeUAV}
Y.~Pan, R.~Li, X.~Da, H.~Hu, M.~Zhang, D.~Zhai, K.~Cumanan, and O.~A. Dobre, ``Cooperative trajectory planning and resource allocation for uav-enabled integrated sensing and communication systems,'' \emph{IEEE Trans. Veh. Technol.}, vol.~73, no.~5, pp. 6502--6516, May 2024.

\bibitem{skolnik1960theoretical}
M.~I. Skolnik, ``Theoretical accuracy of radar measurements,'' \emph{IRE Trans. Aeronaut. Navigational Electron.}, no.~4, pp. 123--129, Dec. 1960.

\bibitem{liu2020radar}
F.~Liu, W.~Yuan, C.~Masouros, and J.~Yuan, ``Radar-assisted predictive beamforming for vehicular links: Communication served by sensing,'' \emph{IEEE Trans. Wirel. Commun.}, vol.~19, no.~11, pp. 7704--7719, Nov. 2020.

\bibitem{CRB}
S.~M. Kay, \emph{Fundamentals of Statistical Signal Processing: Estimation Theory}.\hskip 1em plus 0.5em minus 0.4em\relax Englewood Cliffs, NJ, USA: Prentice-Hall, 1993.

\bibitem{Posterior_1998}
P.~Tichavsky, C.~Muravchik, and A.~Nehorai, ``Posterior {Cramer-Rao} bounds for discrete-time nonlinear filtering,'' \emph{IEEE Trans. Signal Process.}, vol.~46, no.~5, pp. 1386--1396, May 1998.

\bibitem{Beyond_2004}
B.~Ristic, S.~Arulampalam, and N.~Gordon, \emph{Beyond the {Kalman} filter: Particle filters for tracking applications}.\hskip 1em plus 0.5em minus 0.4em\relax Norwell, MA: Artech house, 2003.

\bibitem{AO}
P.~Stoica and Y.~Selen, ``Cyclic minimizers, majorization techniques, and the expectation-maximization algorithm: a refresher,'' \emph{IEEE Signal Process. Mag.}, vol.~21, no.~1, pp. 112--114, Jan. 2004.

\bibitem{CVX}
\BIBentryALTinterwordspacing
S.~B.~M. Grant. {CVX}: Matlab {S}oftware for {D}isciplined {C}onvex {P}rogramming, {V}ersion 2.2. 2020. [Online]. Available: \url{https://cvxr.com/cvx}
\BIBentrySTDinterwordspacing

\bibitem{convex}
S.~P. Boyd and L.~Vandenberghe, \emph{Convex optimization}.\hskip 1em plus 0.5em minus 0.4em\relax New York, NY, USA: Cambridge Univ. Press, 2004.

\end{thebibliography}

\vfill

\end{CJK} 
\end{document}